\documentclass[longauth]{aa}  

\usepackage{graphicx}
\usepackage{txfonts}
\usepackage[
hidelinks,
colorlinks=true,
linkcolor=blue,
citecolor=blue]{hyperref}

\usepackage{graphicx}   
\usepackage{amsmath}

\usepackage{adjustbox}
\usepackage{caption} 
\usepackage[para,online,flushleft]{threeparttable}
\usepackage{rotating} 
\usepackage{booktabs} 
\usepackage{url} 
\usepackage[table]{xcolor}
\usepackage{bm}

\usepackage{longtable}
\usepackage{pdflscape}
\usepackage{gensymb}

\begin{document}

\title{The ALMA survey to Resolve exoKuiper belt Substructures (ARKS)}
\subtitle{I: Motivation, sample, data reduction, and results overview}

\author{S.~Marino\textsuperscript{1}\fnmsep\thanks{E-mail: s.marino-estay@exeter.ac.uk} \and L.~Matr\`a\textsuperscript{2} \and A.~M.~Hughes\textsuperscript{3} \and J.~Ehrhardt\textsuperscript{4} \and G.~M.~Kennedy\textsuperscript{5,6} \and C.~del~Burgo\textsuperscript{7,8} \and A.~Brennan\textsuperscript{2} \and Y.~Han\textsuperscript{9} \and M.~R.~Jankovic\textsuperscript{10} \and J.~B.~Lovell\textsuperscript{11} \and S.~Mac~Manamon\textsuperscript{2} \and J.~Milli\textsuperscript{12} \and P.~Weber\textsuperscript{13,14,15} \and B.~Zawadzki\textsuperscript{3} \and R.~Bendahan-West\textsuperscript{1} \and A.~Fehr\textsuperscript{11} \and E.~Mansell\textsuperscript{3} \and J.~Olofsson\textsuperscript{4} \and T.~D.~Pearce\textsuperscript{6} \and A.~Bayo\textsuperscript{4} \and B.~C.~Matthews\textsuperscript{16,17} \and T.~L\"ohne\textsuperscript{18} \and M.~C.~Wyatt\textsuperscript{19} \and P.~\'Abrah\'am\textsuperscript{20,21,22} \and M.~Bonduelle\textsuperscript{12} \and M.~Booth\textsuperscript{23} \and G.~Cataldi\textsuperscript{24,25} \and J.~M.~Carpenter\textsuperscript{26} \and E.~Chiang\textsuperscript{27} \and S.~Ertel\textsuperscript{28,29} \and A.~S.~Hales\textsuperscript{30,26,14} \and 
Th.~Henning\textsuperscript{31} \and \'A.~K\'osp\'al\textsuperscript{20,22,31} \and A.~V.~Krivov\textsuperscript{18} \and P.~Luppe\textsuperscript{2} \and M.~A.~MacGregor\textsuperscript{32} \and J.~P.~Marshall\textsuperscript{33} \and A.~Mo\'or\textsuperscript{20} \and S.~P\'erez\textsuperscript{13,14,15} \and A.~A.~Sefilian\textsuperscript{28} \and A.~G.~Sepulveda\textsuperscript{34} \and D.~J.~Wilner\textsuperscript{11}} 

\institute{
Department of Physics and Astronomy, University of Exeter, Stocker Road, Exeter EX4 4QL, UK \and
School of Physics, Trinity College Dublin, the University of Dublin, College Green, Dublin 2, Ireland \and
Department of Astronomy, Van Vleck Observatory, Wesleyan University, 96 Foss Hill Dr., Middletown, CT, 06459, USA \and
European Southern Observatory, Karl-Schwarzschild-Strasse 2, 85748 Garching bei M\"unchen, Germany \and
Malaghan Institute of Medical Research, Gate 7, Victoria University, Kelburn Parade, Wellington, New Zealand \and
Department of Physics, University of Warwick, Gibbet Hill Road, Coventry CV4 7AL, UK \and
Instituto de Astrof\'isica de Canarias, Vía L\'actea S/N, La Laguna, E-38200, Tenerife, Spain \and
Departamento de Astrof\'isica, Universidad de La Laguna, La Laguna, E-38200, Tenerife, Spain \and
Division of Geological and Planetary Sciences, California Institute of Technology, 1200 E. California Blvd., Pasadena, CA 91125, USA \and
Institute of Physics Belgrade, University of Belgrade, Pregrevica 118, 11080 Belgrade, Serbia \and
Center for Astrophysics | Harvard \& Smithsonian, 60 Garden St, Cambridge, MA 02138, USA \and
Univ. Grenoble Alpes, CNRS, IPAG, F-38000 Grenoble, France \and
Departamento de Física, Universidad de Santiago de Chile, Av. V\'ictor Jara 3493, Santiago, Chile \and
Millennium Nucleus on Young Exoplanets and their Moons (YEMS), Chile \and
Center for Interdisciplinary Research in Astrophysics Space Exploration (CIRAS), Universidad de Santiago, Chile \and
Herzberg Astronomy \& Astrophysics, National Research Council of Canada, 5071 West Saanich Road, Victoria, BC, V9E 2E9, Canada \and
Department of Physics \& Astronomy, University of Victoria, 3800 Finnerty Rd, Victoria, BC V8P 5C2, Canada \and
Astrophysikalisches Institut und Universit\"atssternwarte, Friedrich-Schiller-Universit\"at Jena, Schillerg\"a{\ss}chen 2-3, 07745 Jena, Germany \and
Institute of Astronomy, University of Cambridge, Madingley Road, Cambridge CB3 0HA, UK \and
Konkoly Observatory, HUN-REN Research Centre for Astronomy and Earth Sciences, MTA Centre of Excellence, Konkoly-Thege Mikl\'os \'ut 15-17, 1121 Budapest, Hungary \and
Institute for Astronomy (IfA), University of Vienna, T\"urkenschanzstrasse 17, A-1180 Vienna, Austria \and
Institute of Physics and Astronomy, ELTE E\"otv\"os Lor\'and University, P\'azm\'any P\'eter s\'et\'any 1/A, 1117 Budapest, Hungary \and
UK Astronomy Technology Centre, Royal Observatory Edinburgh, Blackford Hill, Edinburgh EH9 3HJ, UK \and
National Astronomical Observatory of Japan, Osawa 2-21-1, Mitaka, Tokyo 181-8588, Japan \and
Department of Astronomy, Graduate School of Science, The University of Tokyo, Tokyo 113-0033, Japan \and
Joint ALMA Observatory, Avenida Alonso de C\'ordova 3107, Vitacura 7630355, Santiago, Chile \and
Department of Astronomy, University of California, Berkeley, Berkeley, CA 94720-3411, USA \and
Department of Astronomy and Steward Observatory, The University of Arizona, 933 North Cherry Ave, Tucson, AZ, 85721, USA \and
Large Binocular Telescope Observatory, The University of Arizona, 933 North Cherry Ave, Tucson, AZ, 85721, USA \and
National Radio Astronomy Observatory, 520 Edgemont Road, Charlottesville, VA 22903-2475, United States of America \and
Max-Planck-Insitut f\"ur Astronomie, K\"onigstuhl 17, 69117 Heidelberg, Germany \and
Department of Physics and Astronomy, Johns Hopkins University, 3400 N Charles Street, Baltimore, MD 21218, USA \and
Academia Sinica Institute of Astronomy and Astrophysics, 11F of AS/NTU Astronomy-Mathematics Building, No.1, Sect. 4, Roosevelt Rd, Taipei 106319, Taiwan. \and
The University of Texas School of Law. 727 E. Dean Keeton Street, Austin, Texas 78705, USA
}

\date{Received 18 July 2025; accepted 1 November 2025}

\abstract
 {The outer regions of planetary systems host dusty debris discs analogous to the Kuiper belt (exoKuiper belts), which provide crucial constraints on planet formation and evolution processes. ALMA dust observations have revealed a great diversity in terms of radii, widths, and scale heights. At the same time, ALMA has also shown that some belts contain CO gas, whose origin and implications are still highly uncertain. Most of this progress, however, has been limited by low angular resolution observations that hinder our ability to test existing models and theories. 
   }
   {High-resolution observations of these belts are crucial for understanding the detailed distribution of solids and for constraining the gas distribution and kinematics.
    }
   {We conducted the first ALMA large programme dedicated to debris discs: the ALMA survey to Resolve exoKuiper belt Substructures (ARKS). We selected the 24 most promising belts to best address our main objectives: analysing the detailed radial and vertical structure, and characterising the gas content. The data were reduced and corrected to account for several systematic effects, and then imaged. Using parametric and non-parametric models, we constrained the radial and vertical distribution of dust, as well as the presence of asymmetries. For a subset of six belts with CO gas, we constrained the gas distribution and kinematics. To interpret these observations, we used a wide range of dynamical models. 
   }
   {The first results of ARKS are presented as a series of ten papers. We discovered that up to 33\% of our sample exhibits substructures in the form of multiple dusty rings that may have been inherited from their protoplanetary discs. For highly inclined belts, we found that non-Gaussian vertical distributions are common and could be indicative of multiple dynamical populations. Half of the derived scale heights are small enough to be consistent with self-stirring in low-mass belts ($M_{\rm belt}\leq M_{\rm Neptune}$). We also found that 10 of the 24 belts present asymmetries in the form of density enhancements, eccentricities, or warps. We find that the CO gas is radially broader than the dust, but this could be an effect of optical depth. At least one system shows non-Keplerian kinematics due to strong pressure gradients, which may have triggered a vortex that trapped dust in an arc. Finally, we find evidence that the micron-sized grains may be affected by gas drag in gas-rich systems, pushing the small grains to wider orbits than the large grains. }
   {ARKS has revealed a great diversity of radial and vertical structures in exoKuiper belts that may arise when they are formed in protoplanetary discs or subsequently via interactions with planets and/or gas. We encourage the community to explore the reduced data and data products that we have made public through a dedicated website.}


\maketitle

\section{Introduction}

Young and mature planetary systems contain tenuous dust belts called debris discs, frequently revealed by their infrared excess \citep{Aumann1984}. Due to the dust's short lifetime against radiation and collisional processes, it has long been known that the dust is likely a product of collisions between kilometre-sized or even larger planetesimals \citep{Harper1984, Weissman1984, Backman1993}. Debris discs are therefore the extrasolar analogues of the Solar System's asteroid and Kuiper belts. With ages ranging from tens to thousands of millions of years, debris discs provide a unique window into the final assembly of planetary systems \citep{Wyatt2008, Hughes2018}. Moreover, they allow us to connect the structure seen in protoplanetary discs with the currently known mature exoplanet population. 

At these ages, the gas densities are generally so low that the dynamics are no longer governed by gas, but rather by gravitational interactions (e.g. with planets) and radiation forces. Therefore, debris discs probe a different epoch in the planet formation process: giant planets have already formed, but may still be migrating and scattering material \citep[like Neptune's migration into the Kuiper belt,][]{Fernandez1984, Ida2000}, and terrestrial planets may be in the final stages of formation \citep[e.g. undergoing dust-producing giant impacts like the Earth's Moon-forming impact,][]{Jackson2014, Genda2015}.

Despite the apparent distinction between protoplanetary discs and debris discs, defining which objects are or are not debris discs is not always easy as the evolution from protoplanetary to debris disc is a chaotic process that progresses at different rates throughout the system. Therefore, they share multiple properties \citep{Wyatt2015}. They can both have gas \citep{Kospal2013, Lieman-Sifry2016, Ansdell2016}, and their ages overlap \citep{Lovell2021lupus, Michel2021, Matra2025}. For example, protoplanetary discs can be found at ages as old as 30~Myr \citep{Silverberg2020, Long2025}, and debris discs can be found at ages as young as ${\sim}6$~Myr in the TW~Hydrae association \citep{Miret-Roig2025} and in even younger associations, for example class III sources \citep{Lovell2021lupus}. A working distinction lies in the amount of dust in their discs as measured by the dust fractional luminosity \citep{Hughes2018}. Debris discs typically exhibit fractional luminosities below $1\%$,\footnote{There is a class of debris discs known as extreme debris discs that exhibit fractional luminosities above this threshold and that can vary in time \citep[e.g.][]{Moor2021}}  which makes them generally optically thin at all wavelengths. In contrast, protoplanetary discs have fractional luminosities generally above 1\%, making them optically thick at short wavelengths. In practice, the transition between protoplanetary and debris discs is physical and gradual rather than categorical as systems evolve from being gas-rich and optically thick to dust populations increasingly dominated by second-generation grains produced by planetesimal collisions.

Beyond quantifying the amount of dust in debris discs, resolving its distribution has been fundamental to understanding these systems, starting with the first image of the archetype debris disc $\beta$ Pictoris \citep[in scattered light,][]{Smith1984}. Detailed models provided strong support for the dust-producing planetesimals scenario \citep[e.g.][]{Lecavelier1996}. These models also suggested the potential importance of imaging debris discs at (sub)millimetre wavelengths to trace the parent planetesimals.  The first submillimetre images of debris discs with single-dish telescopes resolved belt-like morphologies at Solar System scales in some very nearby systems \citep[Fomalhaut and $\epsilon$ Eridani,][]{Holland1998, Greaves1998}. Early millimetre interferometry provided sufficient resolution to firmly establish the connection between scattered light and the millimetre belt structures, as predicted by size-dependent dust dynamics \citep[e.g. $\beta$~Pictoris,][]{Wilner2011}.

In the last 10 years, ALMA has revolutionised the study of cold debris discs analogous to the Kuiper belt (exoKuiper belts). Thanks to its unprecedented sensitivity and angular resolution at millimetre wavelengths, ALMA is uniquely suited to reveal the spatial distribution of millimetre-sized grains \citep{Terrill2023}. These large dust grains are mostly unaffected by radiation forces, and thus are ideal tracers of their parent planetesimals and the gravitational interactions in these systems \citep{Krivov2010}. ALMA observations have revealed a wide range of structures, such as narrow belts \citep{Marino2016, Kennedy2018}, wide and smooth belts \citep{Hughes2017, Faramaz2021}, eccentric belts \citep{MacGregor2017, MacGregor2022, Faramaz2019}, belts with gaps \citep{Marino2018, Marino2019, Marino2020hd206, MacGregor2019, Nederlander2021}, belts with possible clumps \citep{Dent2014, Lovell2021, Booth2023}, and a wide diversity of vertical distributions \citep{Daley2019, Matra2019, Hales2022, Han2022, Terrill2023, Marshall2023}. 

Similarly, ALMA high-resolution observations of dust in protoplanetary discs, the predecessors of debris discs, have demonstrated the ubiquity of substructures in the form of rings, gaps, spirals, crescent-shaped features, and warps \citep[see review by][]{Bae2023}. Constraining whether this ubiquity of substructures is passed along to the second-generation dust in exoKuiper belts is fundamental to understanding the evolution of circumstellar discs and planet formation processes. 

In addition, ALMA's unprecedented sensitivity to gas has revealed that several exoKuiper belts contain CO and CI gas that is roughly co-located with the dust \citep[e.g.][]{Kospal2013, Dent2014, Lieman-Sifry2016, Moor2017, Higuchi2017, Cataldi2018}. Most systems with detected gas are young (<50~Myr) A-type stars, but a few exceptional detections have extended this sample to later types \citep{Marino2016, Kral2020, Matra2019twa7} and older systems \citep{Matra2017, Marino2017etacorvi}. The origin of this gas is still unclear. For systems with low CO gas masses, the short photodissociation lifetime of CO implies that CO is continuously replenished via the destruction of volatile-rich solids \citep[i.e. of secondary origin,][]{Marino2016, Kral2016, Matra2017}. Systems with high levels of CO gas are all young enough (10--50~Myr) that the gas could plausibly be part of a protoplanetary disc that has not yet dispersed, with CO molecules shielded by H$_2$ \citep[i.e. of primordial origin,][]{Kospal2013, Nakatani2021}. The high CO gas levels could also be achieved in the secondary scenario since CO is subject to self- and CI-shielding \citep{Kral2019, Marino2020gas}; shielding also explains the non-detection of molecules other than CO in debris discs \citep{klusmeyer2021,smirnov-pinchukov2022}. However, CI shielding is only efficient if vertical mixing is weak, which is currently unconstrained \citep{Marino2022}, and the CI inferred abundance from observations seems insufficient to shield the CO significantly \citep{Cataldi2023, Brennan2024}.

These recent advancements have triggered fundamental questions in our understanding of exoKuiper belts, but most of what we know about dust and gas substructures is based on a handful of the brightest belts. Most belts observed by ALMA have been resolved only enough to determine their widths \citep{Matra2025}. Therefore, it is unclear how common the radial and vertical structures of dust and gas are that have been well resolved in these few systems. Due to the faintness of debris discs, determining the prevalence of these structures over a broader sample requires a significant time investment that only an ALMA large programme can achieve. 

Here we present the first ALMA large programme dedicated to debris discs, The ALMA survey to Resolve exoKuiper belt Substructures (ARKS), to study their detailed dust and gas distribution. This paper is structured as follows. Section~\ref{sec:motivation} describes the motivation and goals of this programme. Section~\ref{sec:strategy} presents the observing strategy and Sect.~\ref{sec:sample} the sample selection. Section~\ref{sec:observations} describes the observations, data reduction, and imaging procedure. Section~\ref{sec:datarelease} describes the data products that we have made available to the community. Section~\ref{sec:results} presents an overview of the results of ARKS presented in nine additional companion papers. Finally, Sect.~\ref{sec:conclusions} summarises the main conclusions of this programme.

\section{ARKS motivation and goals}
\label{sec:motivation}

ARKS was motivated by three core components of exoKuiper belts: the radial distribution of dust, the vertical distribution of dust, and the overall distribution and kinematics of CO gas. Below, we describe the most important findings in these areas that motivated ARKS and our programme design.

The radial and vertical structure of exoKuiper belts had been studied through several ALMA studies on individual systems and a few studies of small samples, until the ALMA survey REsolved ALMA and SMA Observations of Nearby Stars \citep[REASONS,][]{Matra2025}. REASONS compiled and expanded the sample of resolved belts with ALMA and SMA to 74 and performed uniform parametric modelling to constrain their central radius ($R$), full width at half maximum ($\Delta R$), inclination ($i$), position angle (PA), and belt flux ($F_{\rm belt}$). This analysis resulted in a series of important findings for the radial and vertical structure of exoKuiper belts. 

First, REASONS found that most millimetre bright belts are broad discs rather than narrow rings, with fractional widths ($\Delta R/R$) much greater than both the Kuiper belt and rings in protoplanetary discs. This result comes as a surprise, as narrow rings in protoplanetary discs are ubiquitous \citep{Huang2018, Long2019, Cieza2021} and are thought to be the birthplace of planetesimals at tens of au \citep{Stammler2019}. If planetesimals in exoKuiper belts truly formed in these rings, their larger width could be explained by: a) the migration of rings in protoplanetary discs, leaving behind planetesimals and thus producing wide exoKuiper belts \citep{Miller2021}; {b) dynamical perturbations by planets, massive planetesimals or flybys widen these belts \citep[e.g.][]{Booth2009, Lestrade2011}; or c) many broad exoKuiper belts may have unresolved substructures in the form of narrow rings and gaps, as shown for a handful of systems with high-resolution observations \citep{Marino2018, Marino2019, Marino2020hd206, Nederlander2021}.   

Second, REASONS was also able to constrain the vertical thickness or scale height ($H$, vertical standard deviation) for a small subsample of highly inclined belts. The belt vertical aspect ratio ($h=H/R$) is a direct result of the dispersion of orbital inclinations ($i_{\rm rms}$), and thus, constraints on $h$ are translated into constraints on the dynamical excitation of belts. REASONS found $i_{\rm rms} {\sim}1-20\degr$, implying a diversity of stirring levels \citep[see also][]{Terrill2023}. However, these values could be biased by the low resolution of most observations and model assumptions (e.g. Gaussian radial and vertical distributions). Moreover, they could hide more complex vertical distributions, such as the one found in $\beta$~Pic that deviates significantly from Gaussian and suggests at least two dynamical populations \citep{Matra2019}.

ALMA observations of gas have also suffered from similar limitations, with generally low spatial and spectral (velocity) resolution. Spatial resolution is key to assessing how correlated the dust and gas spatial distributions are. In the secondary origin scenario, for instance, CO gas is expected to be co-located with dust unless it is shielded and can viscously spread \citep{Kral2016, Kral2019, Marino2020gas}. There are mixed results when comparing the gas and dust distribution. In some systems, the gas distribution appears to be more extended inwards or outwards \citep{Kospal2013, Moor2013}. In others, the gas distribution appears to be less extended than that of the dust \citep{Hughes2017, Higuchi2019, Hales2022}, and in some cases, they are consistent with each other \citep{Marino2016, Matra2017betapic, Matra2019}. These results are, however, limited by low spatial resolution, hindering a detailed and model-free comparison between the gas and dust \citep[despite the high surface brightness of CO line emission,][]{Moor2017}. The low spectral resolution has also hindered kinematic studies of the gas, which can constrain its kinetic temperature, Keplerian deviations due to planets \citep{Perez2015}, strong pressure gradients \citep{Teague2018}, and outflows \citep{Lovell2021nolup}. It is yet to be seen whether such kinematic features are common among gas-rich exoKuiper belts.

ARKS aims to expand our understanding of debris discs' radial and vertical dust structures, as well as gas distribution and kinematics by performing an ALMA high-resolution survey of 24 belts (new observations of 18 targets and archival data of 6; see Sect.~\ref{sec:sample} for details). In particular, ARKS has the following goals:

\begin{itemize}
\item Dust radial structure: determine what types of dust substructures such as rings/gaps are present in wide belts, down to the level of $\Delta R/R \sim 0.2$ that corresponds to the median of the ring width distribution for bright protoplanetary discs \citep{Matra2025}. If wide belts have narrow rings similar to protoplanetary discs \citep{Bae2023}, this would suggest that planetesimal belts inherit some of the dust structures. Conversely, smooth and featureless belts may, for instance, favour a ring migration or scattered-disc scenarios \citep{Miller2021, Geiler2019}. Gaps could also indicate the presence of planets embedded in these discs, clearing their orbits from debris \citep{Marino2018, Friebe2022}, or closer-in opening gaps through secular interactions \citep{Pearce2015, Yelverton2018, Sefilian2021, Sefilian2023}. Finally, the sharpness of inner and outer edges can inform us about the collisional evolution and dynamics of these belts \citep{Marino2021, Rafikov2023, Imaz-Blanco2023, Pearce2024}.

\item Dust vertical structure: determine the scale heights and vertical dust distributions of highly inclined belts, thus constraining their dynamical excitation level. The excitation level can be used to place constraints on the mass and number of bodies stirring the belts \citep{Ida1993, Quillen2007,Daley2019,Matra2019}. In addition, we can also search for multiple dynamical populations as found in $\beta$~Pic \citep{Matra2019}. This can provide insights into the birth conditions of planetesimals and their interactions with planets, including scenarios involving Neptune-like migration through the belt \citep{Malhotra1995, Matra2019, Sefilian2025}.

\item Gas distribution and kinematics: determine the spatial extent and kinematics of the previously known CO gas in six belts in our sample. By assessing the gas spatial extent relative to dust, we can test whether viscous spreading has taken place \citep[if gas is secondary,][]{Kral2016, Marino2020gas} and whether gas plays a role in shaping the dust distribution \citep{Takeuchi2001, Krivov2009}. The velocity information can also constrain the dynamics of the gas, kinetic temperature, the stellar mass, and the presence of planets.

\item Dust and gas asymmetries: determine if asymmetries are common among the brightest exoKuiper belts at millimetre wavelengths. Asymmetries come in different forms, for example, belt eccentricities \citep{Kalas2005, MacGregor2017, MacGregor2022, Faramaz2019}, density enhancements \citep{Dent2014, Lovell2021, Booth2023}, and warps \citep{Matra2019, Hales2022}. These asymmetries encode information about the dynamical history of these systems, and they are typically interpreted as evidence of interactions with massive companions \citep[e.g.][]{Wyatt1999, Pearce2014, Sefilian2025}.

\end{itemize}

\section{Observing strategy}
\label{sec:strategy}

We defined the following observing strategy to study the radial and vertical distribution of dust and the distribution and kinematics of CO gas.

To study the radial structure of wide belts and determine if they are made of multiple radial components, we first focus on belts with a fractional width larger than 0.3 and that are only moderately inclined, here defined as having an inclination ($i$) lower than $75\degr$, to ease the extraction of radial information.\footnote{Radial information can also be extracted from edge-on belts by performing a deconvolution in the image or visibility space \citep{Han2022, Terrill2023}} The spatial resolution was set with two considerations. First, the beam should be at least as small as the gap width that a Neptune-mass planet would carve in a massless disc if placed on an orbit at the belt central radius $R$ \citep[${\sim}$20\% of $R$ for a stellar mass of $\approx 1 M_{\odot}$,][]{Morrison2015} to search for Neptune analogues. Second, the beam should be small enough to resolve $\Delta R$ into at least five resolution elements to study the sharpness of the inner and outer edges. Finally, we requested a sensitivity such that, after deprojecting and azimuthally averaging the emission, we can measure its radial profile with an average signal-to-noise ratio (S/N$_{r}$) of ${\sim}10$ per resolution element. This strategy meant that each disc was observed with a tailored resolution ($\theta$) and sensitivity (or rms per beam) set by the following:\footnote{This equation was derived by assuming a face-on disc, defining the belt average surface brightness within $\Delta R$ as $I_\nu=kF_{\rm belt}/(2\pi R \Delta R)$, the beam area as $\Omega_{\rm b}=\pi\theta^2/(4\ln(2))$, the signal to noise ${\rm S/N}_r=I_\nu \Omega/{\rm rms}_{\rm av}$, and the rms after azimuthally averaging ${\rm rms}_{\rm av}={\rm rms}\sqrt{\theta/(2\pi R)}$. The factor $k$ was found numerically to be 0.8, assuming a Gaussian distribution. The factor $2\pi R/\theta$ in the rms after azimuthally averaging represents the number of independent points being averaged at radius $R$. We note that for more inclined discs, this number will be lower, but this will be roughly counterbalanced by the higher surface brightness.   }
\begin{flalign}    
    & {\rm rms} = 0.37 \frac{F_{\rm belt}}{ \mathrm{S/N}_{r}}\frac{\theta^{3/2}}{\Delta R \sqrt{R}}, & \label{eq:rms_fo} \\
    & \theta = 0.2\min(\Delta R, R). &
\end{flalign}

To study the vertical structure, we focus on highly inclined belts (here defined as $i>75\degr$) since those are easier to resolve vertically. We aim to resolve the FWHM of the vertical distribution with two resolution elements at the belt central radius $R$, assuming a vertical aspect ratio ($h=H/r$ or vertical standard deviation) of 0.05, and with a S/N of 20 on $h$ (S/N$_{h}$). This would allow us to measure an inclination dispersion ($i_{\rm rms}=\sqrt{2}h$) as low as 1\degr. The required sensitivity to achieve this was found by simulating observations and forward-modelling them, arriving at the empirical relation: 
\begin{flalign}
     & {\rm rms} = 0.63 \frac{F_{\rm belt}}{ \mathrm{S/N}_{h}} \frac{\sqrt{h} \theta^{3/2}}{\sqrt{R} (R+\Delta R/2)}, &  \label{eq:rms_eo} \\
     & \theta = 1.18 hR,  &
\end{flalign}
where we assume $h=0.05$ for our sample, a value found for belts that have been marginally resolved vertically \citep{Terrill2023, Matra2025}.

It should be noted that the constraints on $i$ were in place only when choosing the best targets for observation and constraining the radial and vertical distribution of dust. Nevertheless, we still extract the dust radial distribution of highly inclined belts \citep{rad_arks} and the dust vertical distribution for some belts with inclinations below $75\degr$ \citep{ver_arks}.

We chose to use band 7 for ARKS observations to optimise the required observation time. Band 7 offers the best S/N for a fixed integration time given the typical spectral index/slope of exoKuiper belts \citep[${\sim}2.5$,][]{Matra2025}. However, some systems had already good enough archival band 6 observations, in which case we analysed those (see Sect.~\ref{sec:sample}).

Finally, to study the gas, we targeted the $^{12}$CO and $^{13}$CO~J=3-2 lines in parallel to the continuum, and thus, with the same spatial resolution and integration times. For systems with previously known abundant CO gas, we used the highest spectral resolution available of 26~m/s for $^{12}$CO to extract as much kinematic information as possible. For $^{13}$CO (and $^{12}$CO for targets without abundant gas), we set a lower spectral resolution of 0.9\,km/s to maximise the bandwidth and hence the continuum sensitivity.

\section{Sample selection}
\label{sec:sample}

\begin{figure}
    \centering
    \includegraphics[trim=0.0cm 0.0cm 0.0cm 0.0cm,
    clip=true, width=1.0\columnwidth]{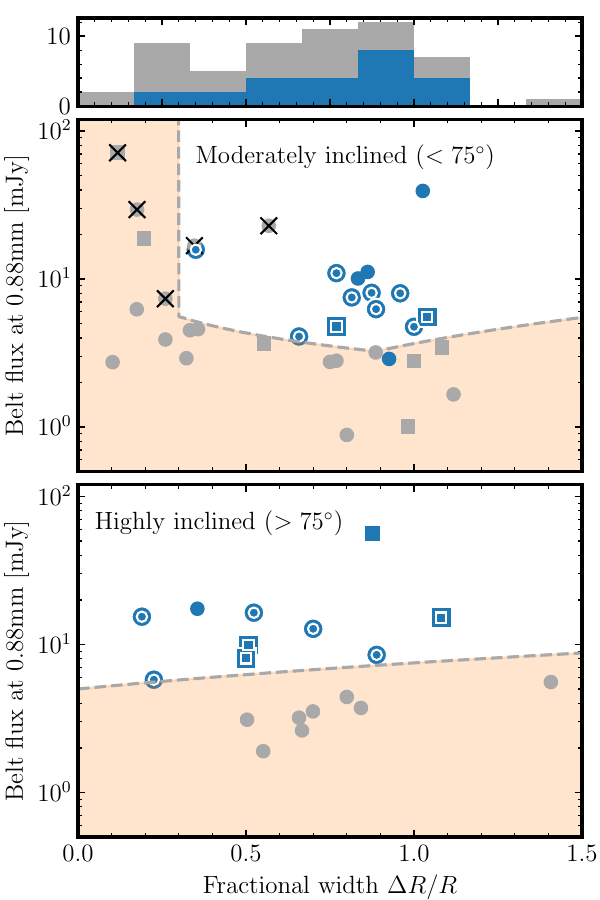}
    \caption{Distribution of fractional widths vs belt fluxes at 0.88~mm. \textbf{Top:} Histogram of all REASONS belts with resolved widths (grey) and those analysed in ARKS (blue, including  six with archival observations). \textbf{Middle:} Distribution of fractional widths vs expected belt fluxes for moderately inclined belts. \textbf{Bottom:} Distribution of fractional widths vs expected belt fluxes for highly inclined belts. The blue markers represent ARKS targets (double markers: newly observed; single markers: archival data). The square symbols represent systems with CO gas, and the black crosses represent systems too large to be observed without mosaicking. The dashed line represents the 10 $\mu$Jy sensitivity limit chosen for ARKS (Eq. \ref{eq:rms_fo} and \ref{eq:rms_eo}), below which belts were excluded unless already observed (orange shaded region). The grey markers represent belts in REASONS that were excluded for being too large, too wide, or too faint for high-resolution observations.}
    \label{fig:sample_selection}
\end{figure}

\begin{figure}
    \centering
    \includegraphics[trim=0.0cm 0.0cm 0.0cm 0.0cm,
    clip=true, width=1.0\columnwidth]{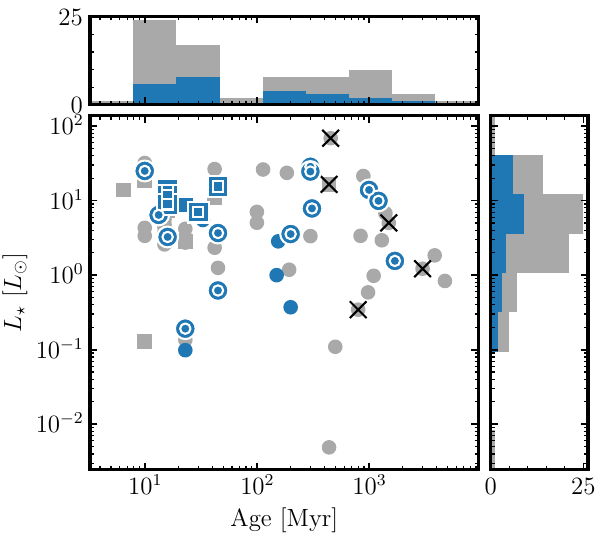}
    \caption{Distribution of stellar luminosities and ages for all systems in REASONS (grey) and in ARKS (blue). The symbols follow the same convention as in Fig.~\ref{fig:sample_selection}. The top and right panels show histograms of the stellar age and luminosities, respectively.  }
    \label{fig:Lstar_vs_age}
\end{figure}

Given the above observing strategy, we selected the most suitable targets from the REASONS sample as those that require the shortest integration times. REASONS comprises nearly all debris discs observed by ALMA with publicly available observations from different programmes, which inherently resulted in a biased sample of the brightest discs, observed in a heterogeneous way. Nevertheless, all these observations were modelled uniformly by fitting a belt model with surface and vertical density distributions described as a Gaussian. 

We started our sample selection by removing those belts in REASONS whose outer edge is larger than 9\arcsec, beyond which the sensitivity drops below 50\% due to the 12m antennas' primary beam in band 7. Smaller belts can be efficiently observed with a single pointing instead of mosaicking. In addition, we removed HD~38858 and HD~36546. The first was only marginally detected with ALMA, and a background source likely contaminated its derived parameters. The second was not well fit by a Gaussian belt model, and thus its derived radius and width are heavily biased by the unsuitable model choice.

To estimate the required sensitivity and thus the integration time of the remaining sample, we used the radius, width, and inclination information from REASONS and the measured and predicted fluxes at 0.88~mm using the existing millimetre fluxes measured between 0.86--1.36~mm and assuming a spectral index of 2.5.  Finally, we focus only on belts with constrained inclinations whose widths have been marginally resolved with an error smaller than 50\%. These considerations left us with 53/74 belts from which we selected our targets.

To obtain a sufficiently large sample, we set the minimum requested rms to 10~$\mu$Jy. This resulted in a complete sample of 24 belts: 14 moderately inclined belts (0--75\degree) and ten highly inclined belts ($>75\degree$), six of which have previously detected CO gas. Six of the 24 systems had archival observations that already met our observation requirements for our goals (Sect.~\ref{sec:strategy}),\footnote{One of the systems with sufficient archival observations (HD~206893) had a flux below our threshold (Fig.~\ref{fig:sample_selection} top panel), but its observations met our resolution and S/N$_r$ criterium.} and thus we did not request new observations. The properties of this sample are summarised in Table~\ref{tab:sample}.

\begin{table*}
    \caption{Main system parameters of the ARKS sample.  \label{tab:sample}
    }
    \centering
    \begin{adjustbox}{max width=0.98\textwidth}
    \begin{threeparttable}
    \begin{tabular}{ l r  c c c c c c | c c c c c }
    \hline
    \hline 
               &      &     &             &             &      &           &                  & \multicolumn{5}{c}{Expected from REASONS} \\
       Name  & Dist & Spt & $L_{\star}$ & $M_{\star}$ & Age  & $L_{\rm cold}/L_{\star}$ & $a_{\rm p}$ & $F_{\rm belt}$ & $i$ & PA & R & $\Delta R/R$  \\
               &  [pc] &   & [$L_{\odot}$] &[$M_{\odot}$] & [Myr] &   [$\times10^{-4}$] & [au]       &         [mJy]        & [$\degr$] & [$\degr$] & [au] & \\
               
    \hline
\multicolumn{10}{l}{Moderately inclined belts ($i<75\degr$)} \\
HD 15257 & 49.0 & F1V & 14 & 1.75  & $1000^{+800}_{-800}$ & 0.61 &  & 7.5 & 40 & 60 & 270 & 0.81 \\
HD 76582 & 48.9 & A7V & 9.9 & 1.61  & $1200^{+900}_{-900}$ & 2.0 &  & 8.0 & 72 & 104 & 219 & 0.96 \\
HD 84870 & 88.8 & A9V & 7.9 & 1.66  & $300^{+200}_{-200}$ & 4.3 &  & 4.8 & 50 & 10 & 260 & 1.00 \\
\rowcolor{gray!10} HD 92945 & 21.5 & K0V & 0.37 & 0.85  & $200^{+100}_{-100}$ & 6.6 & PMa: 2--30 & 10.1 & 67 & 100 & 96 & 0.83 \\
HD 95086 & 86.5 & A9V & 6.4 & 1.54  & $13.3^{+1}_{-0.6}$ & 16 & 52(1) & 8.1 & 29 & 93 & 206 & 0.87 \\
\rowcolor{gray!10} HD 107146 & 27.5 & G0V & 1.00 & 1.04  & $150^{+100}_{-50}$ & 9.8 & PMa: 2.5--20 & 39.3 & 22 & 149 & 107 & 1.03 \\
HD 121617$^{\rm g}$ & 118 & A1V & 14 & 1.90  & $16^{+2}_{-2}$ & 51 &  & 4.8 & 37 & 60 & 78 & 0.77 \\
HD 131835$^{\rm g}$ & 130 & A8V & 8.9 & 1.70  & $16^{+2}_{-2}$ & 23 &  & 5.5 & 74 & 60 & 84 & 1.04 \\
HD 145560 & 121 & F5V & 3.3 & 1.35  & $16^{+2}_{-2}$ & 20 &  & 4.1 & 47 & 28 & 76 & 0.66 \\
HD 161868 & 29.7 & A1V & 24 & 2.11  & $300^{+200}_{-200}$ & 0.65 & PMa: 3--25 & 6.3 & 68 & 57 & 124 & 0.89 \\
HD 170773 & 36.9 & F4V & 3.6 & 1.40  & $200^{+100}_{-100}$ & 4.9 &  & 15.8 & 33 & 114 & 194 & 0.35 \\
\rowcolor{gray!10} HD 206893 & 40.8 & F5V & 2.8 & 1.33  & $160^{+20}_{-20}$ & 2.7 & 3.5,9.7(2) & 2.9 & 24 & 60 & 108 & 0.93 \\
TYC 9340-437-1 & 36.7 & K7V & 0.19 & 0.75  & $23^{+3}_{-3}$ & 10 &  & 11.0 & 40 & 130 & 130 & 0.77 \\
\rowcolor{gray!10} HD 218396 (HR 8799) & 40.9 & A9V & 5.5 & 1.50  & $33^{+7}_{-10}$ & 2.5 & 15,27,41,72(3) & 11.2 & 39 & 51 & 290 & 0.86 \\
\hline
\multicolumn{10}{l}{Highly inclined belts ($i>75\degr$)} \\
HD 9672 (49 Ceti)$^{\rm g}$ & 57.2 & A2V & 15 & 2.00  & $45^{+5}_{-5}$ & 7.2 &  & 15.2 & 79 & 107 & 136 & 1.08 \\
HD 10647 (q$^{1}$ Eri) & 17.3 & F8V & 1.6 & 1.12  & $1700^{+600}_{-600}$ & 2.6 & 2.0(4) & 12.8 & 77 & 57 & 100 & 0.70 \\
HD 14055 & 35.7 & A1V & 29 & 2.19  & $300^{+200}_{-200}$ & 0.89 &  & 8.5 & 81 & 163 & 180 & 0.89 \\
HD 15115 & 48.8 & F3V & 3.7 & 1.43  & $45^{+5}_{-5}$ & 4.7 &  & 5.8 & 88 & 98 & 93 & 0.23 \\
HD 32297$^{\rm g}$ & 130 & A8V & 7.0 & 1.57  & $30^{+10}_{-10}$ & 61 &  & 9.9 & 87 & 48 & 122 & 0.51 \\
\rowcolor{gray!10} HD 39060 ($\beta$ Pic)$^{\rm g}$ & 19.6 & A4V & 8.7 & 1.72  & $23^{+3}_{-3}$ & 25 & 2.7,10(5) & 56.2 & 87 & 30 & 105 & 0.88 \\
HD 61005 & 36.5 & G6V & 0.62 & 0.95  & $45^{+5}_{-5}$ & 25 &  & 16.4 & 86 & 70 & 73 & 0.52 \\
HD 109573 (HR 4796) & 70.8 & B9.5V & 25 & 2.14  & $10^{+3}_{-3}$ & 42 & PMa: 4--38 & 15.4 & 76 & 27 & 78 & 0.19 \\
HD 131488$^{\rm g}$ & 152 & A3V & 12 & 1.80  & $16^{+2}_{-2}$ & 26 &  & 8.1 & 84 & 96 & 92 & 0.50 \\
\rowcolor{gray!10} HD 197481 (AU Mic) & 9.71 & M1V & 0.098 & 0.61  & $23^{+3}_{-3}$ & 3.9 & 0.070,0.12(6) & 17.4 & 88 & 129 & 35 & 0.36 \\
\hline

    \end{tabular}
    \tablefoot{Column 2 indicates the distance according to Gaia DR3 \citep{gaiadr3}. The approximate spectral types are shown in Column 3, estimated using the stellar effective temperatures of REASONS and the main sequence from \cite{Pecaut2013}. The stellar luminosity and fractional luminosity of the cold dust component (Columns 4 and 7) were taken from spectral energy distribution fits using the same methodology as in REASONS \citep{Matra2025}, but with updated disc fluxes based on ARKS (see Appendix \ref{app:fluxes}). The stellar age (Column 6) was taken from REASONS \citep{Matra2025}. The stellar masses in Column 5 were inferred as described in Appendix \ref{sec:star}. Column 8 shows the semi-major axes of known planets in these systems. Systems with unknown planets but with a significant ($>3\sigma$) proper motion anomaly are labelled as PMa \citep{Kervella2022}, with the values originating from \cite{scat_arks}. The last five columns show the expected belt flux at 0.88~mm, the belt inclination, position angle, central radius and fractional width from REASONS. Systems with archival data only and not observed by ARKS are highlighted in grey, and systems with CO gas are flagged  `$^{\rm g}$'.}

    \tablebib{ (1): \cite{Chauvin2018}; (2): \cite{Hinkley2023}; (3): \cite{Zurlo2022}; (4): \cite{Marmier2013}; (5): \cite{Brandt2021betapic}; (6): \cite{Mallorquin2024}.}

    \end{threeparttable}
    \end{adjustbox}

\end{table*}

Figure~\ref{fig:sample_selection} shows the distribution of fractional widths (top panel) vs the expected belt fluxes at 0.88~mm for moderately inclined (middle panel) and highly inclined belts (bottom panel). Blue markers show systems in ARKS, while grey markers represent systems in REASONS that were excluded from ARKS. 

Our sample is biased towards the brightest belts ($\gtrsim3$ mJy). However, as shown in Fig.~\ref{fig:Lstar_vs_age}, the selected sample (blue) is not significantly biased relative to REASONS in terms of stellar luminosities and ages. That said, both samples favour early-type stars with $L_{\star}>1\ L_{\odot}$ and display a bimodal age distribution, with a notable gap between 50 and 100 Myr. The luminosity bias arises because early-type stars are more likely to host bright exoKuiper belts, which are easier to detect and resolve with ALMA. The gap in the age distribution, on the other hand, reflects the absence of moving groups within the 50--100 Myr range in the REASONS sample, resulting in an apparent bimodality \citep{Matra2025}. Of the nearby moving groups, only the recently discovered Volans-Carina falls within this age interval \citep[as it is ${\sim}90$ Myr old,][]{Gagne2018}. Overall, the ARKS sample does not show significant biases relative to REASONS in terms of fractional widths, stellar luminosities, or ages.

In terms of companions, 8 out of the 24 systems have confirmed companions with a wide range of mass ratios and separations (see Appendix~\ref{app:companions} for details). Four of these eight systems (HD~10647, HD~76582, HD~109573 and HD~197481) have stellar companions at wide separations and at least 6 times further than their belts. One of these eight systems has a brown dwarf companion (HD~206893) interior to the belt. Six of the eight systems (HD~10647, HD~39060, HD~95086, HD~197481, HD~206893 and HD~218396) contain companions in the planetary mass regime interior to their belts, at separations that are not greater than 50\% of the belts' inner edge. \cite{scat_arks} presents a summary of the known substellar mass companions in the sample, and their location relative to the belts. In addition, \cite{scat_arks} presents mass upper limits based on SPHERE direct imaging observations, that constrain the possible mass of companions at the disc inner edge (and gaps if present), and semi-major axes and mass constraints for companions causing the PMa in four systems where these have not been confirmed yet via direct imaging.

\section{Observations, data reduction, and imaging}
\label{sec:observations}

In this section we summarise the observations of ARKS and those archival observations that we included.

\subsection{Observations}

ARKS observations (2022.1.00338.L, PI: S. Marino, co-PIs: A. M. Hughes \& L. Matr\`a) were carried out during cycles 9 and 10 from October 2022 to July 2024. Table \ref{tab:obs} summarises the observations of the 18 ARKS and 6 archival targets. For two of the ARKS targets (HD~9672 and HD~10647), we included archival band 7 observations to improve their S/N. The last column of Table~\ref{tab:obs} indicates the project code of the archival observations that were included.

ARKS observations used a wide range of antenna configurations, from the 7m Atacama Compact Array (ACA, with baselines as short as 8m) for the largest belts, to configuration C-8 of the 12m array (with baselines as long as 8.5~km) for the smallest belts. These provided a wide range of resolutions tailored to each system, from 0\farcs85 down to 0\farcs04.

ARKS observations used two different spectral setups depending on the CO content of each system, which had been established by previous observations. For gas-poor systems (i.e. no detected CO gas), two of the four spectral windows were dedicated to studying the dust continuum emission only, with a bandwidth of 1.875\,GHz and 120 channels each and centred at 343.2 and 333.2 GHz. The other two were set to cover the $^{12}$CO and $^{13}$CO J=3-2 lines to search for gas with a bandwidth of 1.875\,GHz, centred at 345.2 and 331.3 GHz, with 3840 channels and a channel spacing of 488\,kHz, providing a spectral resolution of 0.84 and 0.85~km/s, respectively. The total continuum bandwidth for these systems was 7.5\,GHz.

For systems regarded to be gas-rich (HD~9672, HD~32297, HD~121617, HD~131488, HD~131835), the spectral window covering the $^{12}$CO J=3-2 emission at 345.789\,GHz was set with the highest spectral resolution of 31\,kHz or 26\,m/s (15~kHz or 13\,m/s channel spacing). This spectral window had a total bandwidth of 59\,MHz or 51\,km/s, with 3840 channels and was centred at the line central frequency, taking into account the radial velocity of each system. The total continuum bandwidth for these systems was 5.7\,GHz.

In addition to the new ARKS observations, we include ALMA archival observations for 2 of the 18 systems observed by ARKS. These systems had previous observations at a similar resolution in band 7, although at a lower sensitivity. However, these observations still enhance our S/N due to the increased integration time. These systems are HD~9672 \citep[49~Ceti,][]{Cataldi2023, Delamer2023} and HD~10647 \citep[q$^{1}$~Eri,][]{Lovell2021}. For the 6 additional archival targets, we preferentially use their band 7 data with two exceptions. For HD~206893, its band 6 data has a higher S/N at the desired resolution, and thus we use this only \citep{Marino2020hd206, Nederlander2021}. For HD~107146, we use both bands 6 and 7 since they are complementary: the former has high S/N, low-resolution observations, while the latter includes higher-resolution data at a moderate S/N \citep{Marino2018, Marino2021, Imaz-Blanco2023}. The spectral setup of these observations varied, with most of them including $^{12}$CO observations while maximising the continuum sensitivity. Finally, 5 of the 6 archival targets were observed with a single pointing, the exception being HD~39060 ($\beta$~Pic), which included a 2-point mosaic for its 12m antenna short baseline observations.

\subsection{Data reduction}
\label{sec:reduction}

The data were calibrated by the UK ARC Node using the ALMA pipeline and corresponding CASA version \citep[6.4.1.12 for those observed during cycle 9 and 6.5.4.9 for cycle 10,][]{casa}. For archival data, the calibrated measurement set (MS) files were delivered by the ESO ARC. No additional steps were done for the calibration of the data, except for some flagging of some archival data (see below).

We reduced the calibrated MS files using CASA version 6.4.1.12 as follows. We first transformed the MS files to the barycentric reference frame using the task \textsc{mstransform} and kept only the target observations. We then time averaged the data using the task \textsc{split}. The time binning ($\Delta t$) was calculated for each source and antenna configuration such that the effect of time averaging smearing is kept below 5\% at a radius equal to twice the size of the belts' outer edges (following Eq. 6.80 in \citealt{isra}). We also constrained the maximum value of $\Delta t$ to 60~s, which was the case for most of our sources. For HD~61005 and $\beta$~Pic long baseline observations, $\Delta t$ was set to 45 and 30~s, respectively, to keep the effect of time averaging smearing below 5\%. 

Each target observation was centred at the predicted stellar position using the available stellar astrometry in cycle 9. For the targets observed as part of ARKS, this was done using Gaia DR2 \citep{gaiadr2}, except for HD~14055 and HD~161868 for which we used the Hipparcos astrometric solution \citep{hipparcos}. These two stars have the largest apparent magnitudes, which makes their Gaia DR2 astrometric solution less accurate than Hipparcos. We find that the DR2 positions match well with the DR3 positions, with differences smaller than 5 mas. The Hipparcos predicted positions of HD~14055 and HD~161868, and the phase centres used in the archival observations, differed by larger values up to 81 mas. The final phase centre coordinates (after the data correction described in Sect.~\ref{sec:correction}) and the offsets relative to Gaia DR3 are summarised in Table \ref{tab:fluxes}.

To study the dust continuum emission, we flagged channels within 15 km/s of the $^{12}$CO and $^{13}$CO J=3-2 lines\footnote{or the J=2-1 lines for archival sources with band 6 observations.} for systems with known gas, and within 50~km/s for systems with no known CO gas. Subsequently, we spectrally averaged the data by $\Delta \nu$, which was calculated for each source and antenna configuration such that the effect of bandwidth smearing is kept below 5\% at a radius equal to twice the size of the belts' outer edges (following Eq. 6.75 in \citealt{isra}). The spectral averaging is also constrained to be at most the bandwidth of any spectral window (${\sim}2$ or 0.06~GHz), which sets $\Delta \nu$ for most of our sources. The exceptions to this were the long baseline observations of HD~9672, HD~15115, HD~32297, HD~61005, HD~109573 and HD~131488 with a frequency averaging of 1~GHz and HD~39060 long baseline observations with a 0.5~GHz averaging. 

To study the CO gas, we split the time-averaged MS files into two new files containing the spectral windows dedicated to $^{12}$CO and $^{13}$CO J=3-2 lines (or $^{12}$CO J=2-1 for HD~39060's archival data). We used the task \textsc{uvcontsub} to fit and subtract the continuum emission, excluding the same channels that were flagged to study the continuum as described above. 

An additional step was taken to reduce the data of HD~107146, HD~10647 and HD~197481 (AU~Mic) that involved flagging. For HD~107146's band 7 12m short baseline observations, we flagged visibilities at baselines shorter than 20~k$\lambda$ to reduce some low-frequency ripples in the image that were not caused by the uv coverage but rather systematic noise in the data. This solves most of the problem, but we note that some large-scale artefacts remain in the data, especially in images constructed with more weight towards the band 7 short baselines \citep{Imaz-Blanco2023}. For HD~10647 long baseline observations in cycle 3 taken on 12 Jul 2016 at 11:34 (UTC), we flagged the baselines corresponding to the antenna pair DA46 and DA54 since they produced strong ripples in the continuum image \citep{Lovell2021}. For HD~197481 (AU~Mic), we flagged scan 27 in the long baseline observations taken on 24 Jun 2015 (UTC) to remove a stellar flare from our data \citep{Daley2019, MacGregor2020}. We also searched for stellar variability on all the data, but found no significant variability for any target other than HD~197481 (see Appendix \ref{sec:variability}). Finally, we note that HD~218396 has background CO J=3-2 emission at -12.5 km/s in the barycentric reference frame \citep{Faramaz2019}, which was flagged as part of our standard reduction.

ExoKuiper belts are not typically bright enough for successful self-calibration. Nevertheless, as an experiment, we tested whether self-calibration could improve our data's S/N. We tried self-calibrating our brightest targets using the package \textsc{auto\_selfcal}\footnote{Available at \href{https://github.com/jjtobin/auto\_selfcal/}{https://github.com/jjtobin/auto\_selfcal/}} but found no improvements in the image rms. Therefore, we continue using the non-self-calibrated data.

\subsection{Data correction}
\label{sec:correction}

Before imaging and analysing the continuum and gas data, we applied a final data correction step, which aims to correct any astrometric offsets, flux offsets, and rescale the visibility weights of each execution block to truly represent the noise or dispersion in the continuum and CO data. For sources with background galaxies, this step also includes subtracting their emission from the continuum data. To determine these corrections, we use the continuum data as a reference.

We start this process by extracting the continuum visibilities of each execution block\footnote{Each system can be observed with multiple antenna configurations (scheduling blocks), and multiple executions per configuration (execution blocks).} and forward-fitting them simultaneously using a simple belt model (see below), leaving as free parameters astrometric (phase centre) offsets, flux scale offsets, and weights scale offsets for each execution block (in addition to system parameters), and exploring the parameter space with an MCMC \citep[as in][]{Marino2021}. This fitting and correction step is effectively a low-order self-calibration using a simple model as a reference.

The belt model was computed using \href{https://github.com/SebaMarino/disc2radmc}{\textsc{disc2radmc}} \citep[a Python wrapper to use RADMC3D\footnote{https://www.ita.uni-heidelberg.de/~dullemond/software/radmc-3d/},][]{Marino2022}. For most systems, we assumed a belt with a Gaussian surface density distribution (with central radius and FWHM as free parameters) and a Gaussian vertical distribution with a free and radius-independent aspect ratio for sufficiently inclined systems (a similar strategy to REASONS). The exceptions to this were HD~15115, HD~92945 and HD~107146, for which we fit a double Gaussian radial distribution with additional free parameters for the central radius, FWHM, and relative scale of the second Gaussian; HD~206893, for which we fit a Gaussian gap with additional free parameters for its centre, width, and depth; and HD~121617, for which we included a sinusoidal azimuthal variation (with additional parameters for the phase and amplitude of the variation). Without these additional components, the best fits to these systems showed strong residuals and, thus, could strongly bias our correction process. For systems where central emission was detected using uncorrected clean images, we kept the stellar flux as a free parameter. We also considered the belt’s inclination and position angle as free parameters. Finally, we left the dust mass as a free parameter to fit the belt flux, assuming a dust opacity of 1.9~cm$^{2}$~g$^{-1}$ at 0.89~mm or 1.3~cm$^{2}$~g$^{-1}$ at 1.3~mm.\footnote{Computed using Mie Theory for a power law grain size distribution ranging from 1~$\mu$m to 1~cm, with a slope of -3.5, and composed of astrosilicates (70\%), crystalline water ice (15\%) and amorphous carbon (15\%) by mass \citep{Draine2003, LiGreenberg1998}.} Since we fit two bands for HD~107146, we included an extra parameter to fit the belt flux in band 6 (1.3~mm) instead of fitting the spectral index or grain size distribution.

In addition, we account for background submillimetre galaxies (SMGs) present near the belts of HD~76582, HD~84870, HD~92945, HD~95086, HD~107146, HD~206893, TYC 9340-437-1, and HD~218396, by including 2D elliptical Gaussians in the model images. We left as free parameters their flux, position relative to the stars, \citep[taking into account the SMG's relative motion to these stars due to the stellar proper motions,][]{gaiadr3}, major and minor axes, and position angle. Fitting these background sources had the advantage of providing an additional lever to determine any systematic astrometry or flux offsets between observations. Appendix~\ref{app:smg} explains why these are likely SMGs and presents the best-fit values for these sources.

The model images were multiplied by the antenna primary beam (computed using CASA) and then Fourier transformed with Galario to produce model visibilities to fit the ALMA data \citep{galario}.
To account for any systematic phase centre offsets between observations, we left as free parameters a right ascension and declination offset for each execution block. This is applied to the model by Galario in Fourier space. 

For each model evaluation, Galario computes a $\chi^2$ for each execution $j$ that is defined as
\begin{equation}
    \chi^2_j(x) =  \sum_{k}^{N_j} ||V_{\mathrm{d},k, j}f_{\mathrm{flux},j} - V_{\mathrm{m}, k,j}(\bm{x})P_{k,j}||^{2} \frac{W_{k,j}}{f_{\mathrm{flux},j}^2 f_{\sigma, j}^2 }, 
    \label{eq:chi2j}
\end{equation} 
where $V_{\mathrm{d},k, j}$ and $V_{\mathrm{m}, k,j}(x)$ are the measured and model complex visibilities $k$ of the execution $j$, $f_{\mathrm{flux},j}$ is the flux correction factor, $\bm{x}$ represents the system's free parameters (including background sources), $P_{k,j}$ represents the phase centre offset with free parameters $\Delta{\rm RA}_{j}$ and $\Delta {\rm Dec}_j$,\footnote{In Eq. (\ref{eq:chi2j}), $P_{k,j}=\exp[-2\pi i (u_{k,j}\Delta {\rm RA}_{j} + v_{k,j} \Delta {\rm Dec})]$, where $u_{k,j}$ and $v_{k,j}$ are the $u-v$ coordinates of each visibility.} $W_{k,j}$ is the associated weight (1/uncertainty$^{2}$), $f_{\sigma, j}$ is a factor to rescale the weights (see below), and $N_j$ is the number of visibilities for a particular execution. Each system was typically observed with two antenna configurations and multiple executions per configuration. Because each execution was observed and calibrated independently, they may have different systematic astrometric offsets.

Similarly, each execution can have different absolute flux calibration errors of the order of ${\sim}$10\% \citep{Francis2020}. To account for this, we fit a flux correction factor $f_{\mathrm{flux}, j}$ to each execution $j>0$ while fixing $f_{\mathrm{flux}, 0}=1$ to anchor these factors. We arbitrarily assigned $j=0$ to one of the executions taken with the longest baseline configuration of each target.\footnote{Although we make sure we do not use as a reference execution an observation with a significantly higher noise than the average.} For HD~107146, we had to set $f_{\mathrm{flux}}=1$ for one band 7 execution and one band 6 execution.

Finally, the factors $f_{\sigma, j}$ in Eq. (\ref{eq:chi2j}) account for the fact that the weights in the calibrated MS files do not necessarily match the true dispersion and uncertainty in the data. While the task \textsc{statwt} in CASA can empirically determine the weights, here we choose to conserve the relative weights between different measured visibilities and only adjust their absolute level \citep[e.g.][]{Marino2018, Matra2019}. We tested whether this offset varies with spectral window and found it did not, and thus, we chose to use a single factor per execution. Fitting for $f_{\sigma, j}$ is equivalent to forcing the reduced $\chi^2_{j}$ to be one, which is an adequate choice for ALMA observations of exoKuiper belts where the S/N per visibility is $\ll1$ \citep[i.e. each value is dominated by noise,][]{Marino2021}. This means that the choice of the model has little influence on the best-fit value of $f_{\sigma, j}$.

We use the Python package \textsc{emcee} \citep{emcee} to fit all these parameters using the  Affine Invariant MCMC Ensemble sampler \citep{Goodman2010} and recover the posterior distribution defined as 
\begin{equation}
     \log P = - \frac{1}{2}\sum_j [\chi^2_j(x) + 2N_j\log(f_{\sigma, j})] + \mathrm{prior}, 
     \label{eq:log_P_prior}
\end{equation}
where the prior is uniform for all parameters except for the flux offsets ($f_{\mathrm{flux, j}}$), for which we assume a Gaussian prior with a standard deviation of $10\%$.\footnote{The second term inside the brackets in Eq. (\ref{eq:log_P_prior}) arises from the normalisation of a Gaussian distribution where the standard deviation has been scaled by $f_{\sigma, j}$, which are free parameters. Thus, they must be included explicitly in the definition of $ \log P $.} Typically, each system requires between 50 and 100 free parameters, depending on the number of executions, and the MCMC was run with 200--400 walkers and for 2000--10,000 iterations, depending on how fast the chains converged. After removing the burn-in phase that typically lasted half the iteration length, we extracted the median and 16$\mathrm{th}$ and 84$\mathrm{th}$ percentiles for each parameter.

We used the medians of the fitted parameters to correct the continuum and CO data.\footnote{The medians approximate very well the values with the highest likelihood with differences within one $\sigma$ in most cases, and combined they produce a high-likelihood fit.}  The medians of $\Delta{\rm RA}_{j}$ and $\Delta {\rm Dec}_{j}$ were used to align each execution with our reference execution ($j=0$). In this way, if there was a true offset in the disc relative to the predicted stellar location, this would be conserved, but only with the accuracy that the single $j=0$ execution provides. We also copy the coordinates of this reference execution to all other executions using the CASA task \textsc{fixplanet}. Typically, these offsets were found to be smaller than the beam size and consistent with the expected systematic astrometric uncertainty (see Sect.~\ref{sec:imaging}). Two systems, HD~14055 and HD~39060, showed offsets with a standard deviation between executions larger than 30\% of the beam size (i.e. 3 times the expected absolute astrometric accuracy of 5--10\% of the beam, \href{https://almascience.eso.org/documents-and-tools/cycle11/alma-technical-handbook}{ALMA Technical Handbook}). For HD~14055, we attribute the large dispersion to the low elevation (${\sim}30\degree$) of its observations due to its declination, which may have increased the phase noise, and to a low S/N per execution. For HD~39060 ($\beta$~Pic), this is due to one of three executions that has a much higher noise. The coordinates of the final phase centre and the date of the reference execution of each system are presented in Table~\ref{tab:fluxes}.

The medians of $f_{\mathrm{flux},j}$ were used to correct for the relative flux offsets. These were found to be typically centred at 1 and have a dispersion of 8\% between executions. Finally, the weights of each execution were corrected by using the medians of $f_{\sigma, j}$. We find $f_{\sigma, j}=1.86-1.88$ for the 12m array and $f_{\sigma, j}=1.65-1.70$ for the 7m array ARKS observations. These factors mean that the visibility uncertainties are typically underestimated by 70-90\%, which would impact any analysis performed using the uncorrected visibility weights. For archival observations, there is a wider range of values from 0.99 to 33.3, likely due to differences in the calibration of weights in earlier ALMA cycles.  

For systems with SMGs, we also produce corrected continuum MS files with those sources subtracted using their model visibilities. For most sources, the SMG subtraction performed well, leaving no significant residuals. The two exceptions were HD~95086 and HD~107146. 
The results of this step, including the best-fit parameters of these sources, are presented in Appendix~\ref{app:smg}. We note that the CO data did not require subtracting the SMGs, as the continuum subtraction step subtracted their emission.

HD~39060 ($\beta$~Pic) observations used a mosaic strategy \citep{Matra2019} for the compact 12m observations: two pointings set along the disc major axis and offset by 5\arcsec\ from the star. We found that our offset parameters for the pointing centred on the SW side (single execution) were significantly different from the nominal phase centre by 0.5\arcsec (${\sim}2$ beams). This could be due to our axisymmetric disc assumption as $\beta$~Pic's disc presents slight asymmetries \citep{Dent2014, Matra2019}. Therefore, for the mosaic observations, we fixed $\Delta{\rm RA}_{j}$ and $\Delta {\rm Dec}_{j}$ to zero.

The inferred central radius, FWHM, disc aspect ratio, and stellar flux for the ARKS sample can be found in Table~\ref{tab:mcmc_results}. Overall, we find good agreement between the derived central radius and FWHM of the dust radial distribution and the estimated values from REASONS in Table~\ref{tab:sample}. However, for some systems, we find significant differences in the FWHM that we attribute to the radial profiles being far from Gaussian distributions \citep[see detailed radial profiles in the companion paper by][]{rad_arks}. This is the case for HD~131835, HD~109573, HD~121617, HD~131488 and HD~32297, which display narrow peaks surrounded by extended low-level emission. This translates into ARKS observations being best fit by narrower Gaussians than the lower-resolution REASONS data that are more (less) sensitive to the wide (narrow) component.   

For those systems in which we fit a stellar flux, we find values consistent (within $3\sigma$) with the fluxes expected from their SEDs. These predicted fluxes are extrapolations of their fitted photometry assuming a Rayleigh-Jeans behaviour (presented in brackets in the eighth column of Table~\ref{tab:mcmc_results}). Three stars, however, present significant differences: HD~9672, HD~10647, and HD~197481. Though we do not detect the star in HD~9672 (49~Ceti), we constrain its flux to be <19~$\mu$Jy, which is significantly lower than the 28~$\mu$Jy expected from its photosphere \citep[as has been found before for other A-type stars,][]{White2018}. For HD~10647 (q$^1$~Eri), we find a stellar flux of $131\pm9$~$\mu$Jy ($\pm16$~$\mu$Jy when considering the absolute flux uncertainty) that is higher than the predicted photospheric emission of 96~$\mu$Jy, but still within 3$\sigma$. A higher flux could arise from an unresolved inner belt \citep{Lovell2021}. For HD~197481 (AU~Mic), we find a significantly higher flux of $245\pm14$~$\mu$Jy ($\pm29$~$\mu$Jy when considering the absolute flux uncertainty) than the expected photospheric emission of 50~$\mu$Jy. HD~197481 is known to have variable non-photospheric emission at mm wavelengths, which may explain this higher flux (\citealt{Daley2019}; see also our Appendix~\ref{sec:variability}).

\subsection{Continuum imaging}
\label{sec:imaging}

Image reconstruction was performed using the CLEAN algorithm implemented in the task \textsc{tclean} in \textsc{CASA} \citep{casa}. We used Briggs weighting using robust parameters\footnote{This parameter can take values from -2 to 2, with higher values resulting in a lower resolution but higher S/N. A value of 0.5 is the standard used for observation planning and the quality assessment of the data.} tailored to each source to find the right balance between resolution and S/N for the different data analyses. For example, the analysis of the radial distribution of dust benefits from higher resolution as one can azimuthally average the emission \citep{rad_arks}. Conversely, the analysis of asymmetries benefits more from a higher S/N than higher resolution \citep{asym_arks}. Typically, we used robust values of 0.5, 1.0 and 2.0, except for a few sources where the S/N was sufficient to reduce the robust value to 0 (e.g. for HD~121617 or HD~32297). We also applied a uvtaper for other sources to boost the S/N per beam and better recover the emission in the reconstructed images (e.g. HD~95086). We used the standard gridder by default, except for sources where we combined 12m and 7m array observations (10/24). For those, we used the mosaic gridder option to account for the different primary beams (this also includes HD39060/$\beta$~Pic observations that included a mosaic). Finally, we also used the multiscale option, setting the scales to roughly 0, 1, 3 and 9 times the size of the robust=0.5 beam to recover better larger-scale structures. 

While cleaning, we manually masked the dirty image, only including regions with positive emission and updated these masks between cleaning cycles to include lower surface brightness regions as imaging artefacts disappeared. We stopped cleaning once the residuals outside the mask appeared like Gaussian noise to visual inspection without large-scale artefacts.

For the particular case of HD~107146, we combine its band 6 and 7 data into a single multi-band image using the `mtmfs' deconvolver with two terms and a reference frequency of 300~GHz (1.0~mm). Combining both bands was advantageous as it incorporated the high-resolution information from the band 7 observations and the high signal-to-noise short baseline observations from the band 6 observations. The resulting image was significantly better in S/N than the band 7-only image while keeping the beam size almost unchanged. We note, however, that the multi-band images suffer slightly from large-scale artefacts due to systematic noise in the band 7 data (see Sect.~\ref{sec:reduction}). For this reason, when searching for asymmetries in this system, we use the band 6 data only \citep{asym_arks}.

\begin{figure*}
    \centering
    \includegraphics[trim=0.0cm 0.0cm 0.0cm 0.0cm,
    clip=true, width=0.82\textwidth]{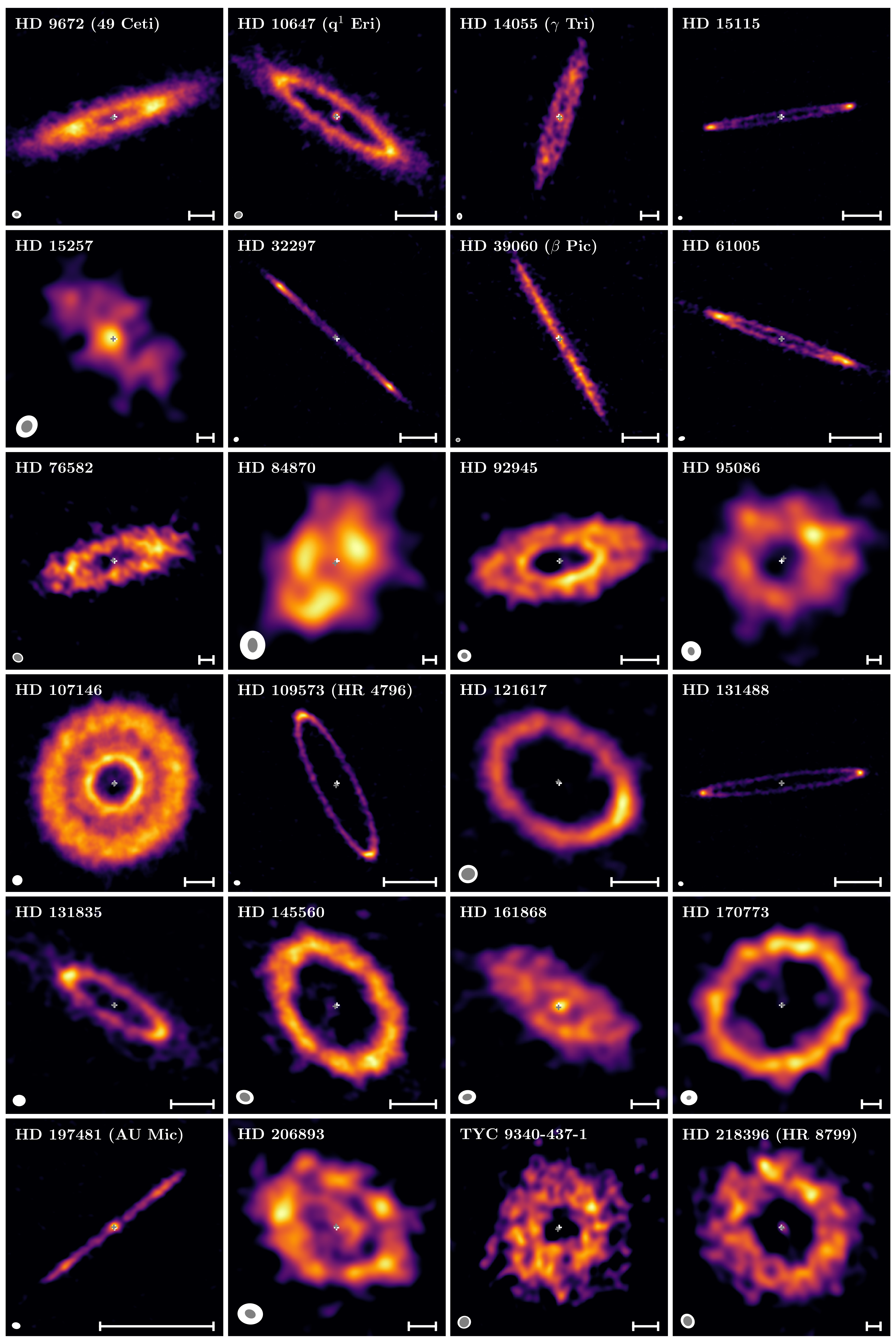}
    \caption{ARKS continuum clean images of the 24 systems in the sample after correction and subtraction of any SMG. The beam size is shown as a white ellipse in the bottom left corner. For sources imaged with a robust parameter greater than 0.5, an additional grey ellipse represents the beam size using a robust value of 0.5. The white ticks at the edges are spaced by 1\arcsec, while the scale bar at the bottom right corner represents a projected distance of 50~au. The white cross represents the expected stellar position according to Gaia DR3, while the grey cross represents the best-fit centre of the system assuming a circular belt. For better clarity, each panel uses its own colour scale from $3\times$rms to the image peak.
     \label{fig:gallery} }
    
\end{figure*}

\begin{figure*}
    \centering
    \includegraphics[trim=0.0cm 0.0cm 0.0cm 0.0cm,
    clip=true, width=1.0\textwidth]{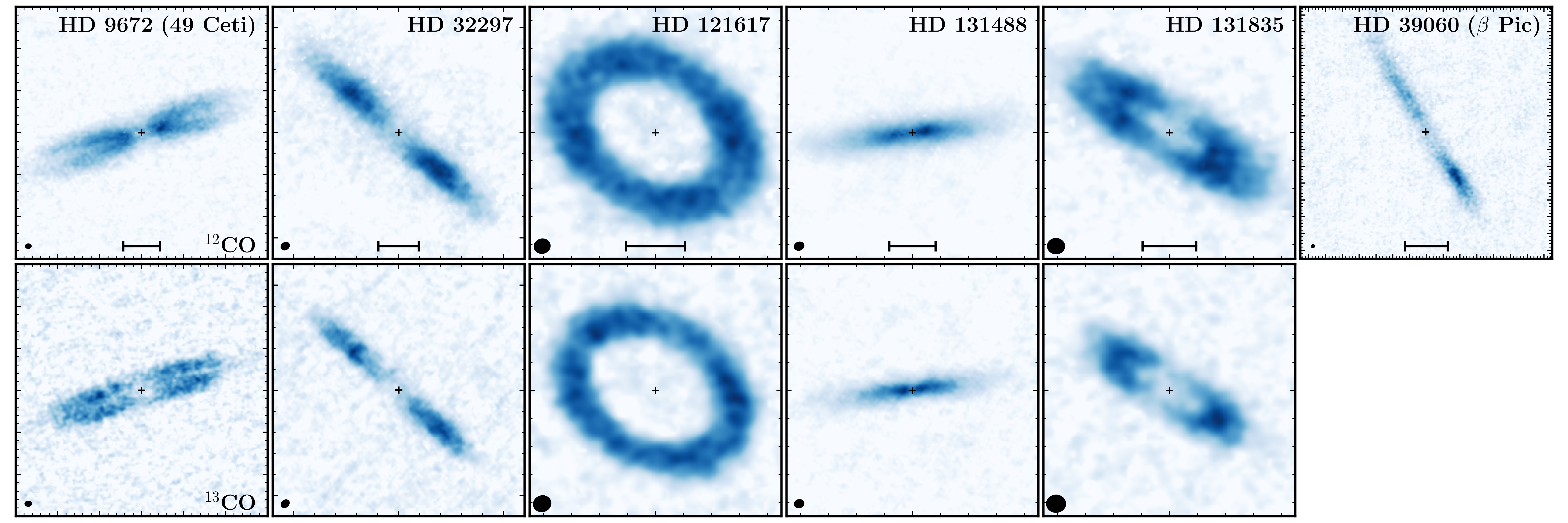}
    \caption{ARKS CO moment 0 images of the six systems with gas in the sample after subtracting the continuum ($^{12}$CO at the top and $^{13}$CO at the bottom row). We  selected an appropriate robust parameter for each system to enhance the S/N per beam if necessary. The beam size is shown as a black ellipse in the bottom left corner. The black cross represents the expected stellar position according to Gaia DR3. The bar at the bottom of the top panels represents 50~au. The large and small ticks are spaced by 1 and 0\farcs2, respectively. The imaging process is described in \cite{gas_arks}. We note that HD~39060 does not have a $^{13}$CO image as it was not included in the archival observations that we use for ARKS.
    }
    \label{fig:gas_gallery_m0}
\end{figure*}

\begin{figure*}
    \centering
    \includegraphics[trim=0.0cm 0.0cm 0.0cm 0.0cm,
    clip=true, width=1.0\textwidth]{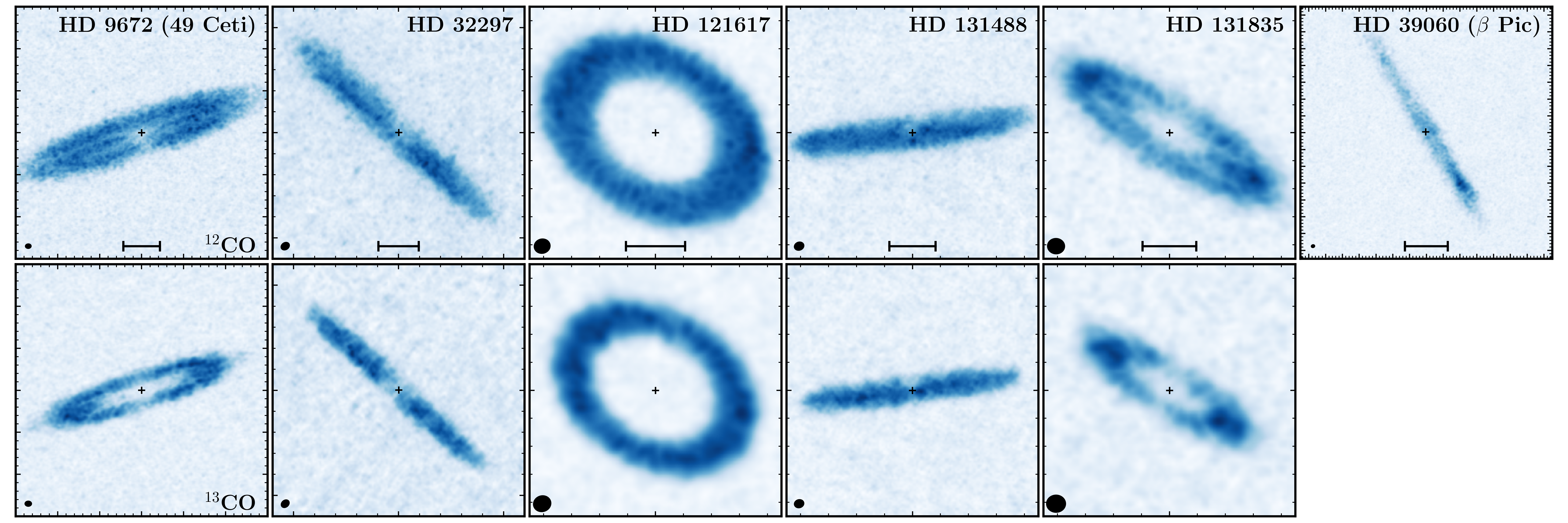}
    \caption{Same as Fig.~\ref{fig:gas_gallery_m0}, but showing the moment 8 (peak intensity) images of the six systems with gas \citep{gas_arks}.
    }
    \label{fig:gas_gallery_m8}
\end{figure*}

Figure~\ref{fig:gallery} presents the continuum images of the SMG-subtracted and corrected data using different robust values to highlight the overall disc emission. The corresponding beam is shown as a white ellipse. The white cross in the image centre shows the expected stellar location according to Gaia DR3 for the ARKS targets \citep{gaiadr3}, while the grey cross shows the best-fit centre from our Gaussian fit. These differences are small but significant\footnote{$>3\sigma$ when looking at the 2D posterior distribution of $\Delta{\rm RA}$ vs $\Delta{\rm Dec}$ for the multiple long baseline observations of each system.} for HD~15115, HD~32297, HD~39060 ($\beta$~Pic), HD~76582, HD~95086, HD~109573 (HR~4796), HD~161868, and HD~197481 (AU~Mic). These offsets may be due to systematic errors (e.g. errors from baseline uncertainties or phase referencing errors) or disc asymmetries (e.g. disc eccentricities or crescents) that bias the Gaussian fit away from the star. Systematic pointing errors are not expected to be larger than 30\% of the beam (3 times the expected absolute astrometric accuracy of 5-10\% of the beam), which would suggest that the offsets in HD~15115, HD~32297, HD~95086 and HD~109573 are due to asymmetric emission. These offsets and the presence of asymmetries are discussed in detail in the companion paper by \cite{asym_arks}.

\subsection{Line imaging}

Line imaging was performed using the CLEAN algorithm and the continuum-subtracted data. We used Keplerian masks tailored to each source, producing line cubes that we used to generate moment maps. The details of this process and gas analysis are described in the companion paper by \cite{gas_arks}. Detailed analysis of the gas distribution, line profile and kinematics of the only source significantly away from edge-on, HD~121617, can be found in \cite{line_arks} and \cite{hd121617_arks}. Figures~\ref{fig:gas_gallery_m0} and \ref{fig:gas_gallery_m8} show the moment 0 and 8 images (integrated intensity and peak intensity, respectively) of the $^{12}$CO and $^{13}$CO J=3-2 emission for the 6 systems with gas detections in ARKS. We note that the HD~39060 band 6 data did not include the $^{13}$CO J=2-1 line. Our observations did not reveal any new, previously unknown, CO-bearing system, likely because the higher resolution achieved in ARKS spread any signal over more resolution elements than prior observations, leading to a disc-integrated detection capability not significantly deeper than prior observations, despite the longer integration times.

\section{Data release}
\label{sec:datarelease}

We have made the corrected data and other data products publicly available on our dedicated website \href{https://arkslp.org}{arkslp.org} and \href{https://dataverse.harvard.edu/dataverse/arkslp/}{ARKS dataverse}. The data products include:
\begin{itemize}
    \item The continuum and CO gas measurement sets (MS files), time averaged, frequency averaged for the continuum, and corrected for astrometric offsets, flux offsets, weights rescaling, and with and without the subtraction of SMGs (see Sect.~\ref{sec:correction}). 
    \item Corrected continuum visibility tables in a text format.
    \item Clean continuum and CO gas images with various resolutions (robust parameter), uv-tapering, and before and after subtracting SMGs.
    \item Reduction and correction scripts to run in CASA to re-generate the data, and best-fit values that we used to correct the data as json files.
    \item A master table with the most relevant information presented in Tables \ref{tab:sample}, \ref{tab:obs} and \ref{tab:mcmc_results}.
\end{itemize}

\section{Initial results overview}
\label{sec:results}

In this section we provide an overview of the results from the first ARKS paper series. 

\subsection{ARKS II: The radial structure of debris discs, \cite{rad_arks}}

In this paper, we analyse the radial distribution of dust using parametric and non-parametric models. We find many structures, such as multiple rings and gaps, halos, and sharp or smooth edges. Overall, we find that 5/24 belts have multiple rings, 7/24 belts have low-amplitude emission (either a halo or additional faint rings), and the remaining 12/24 are consistent with being single belts (some of which have substructures such as shoulders or plateaus). We find a bimodal distribution in the fractional width of rings across the whole sample, with our new observations revealing that the population of narrow rings is more prevalent than the lower-resolution REASONS data suggested. This is not only because several belts are resolved into multiple components, but also because some broad belts have been resolved into a very narrow peak surrounded by extended and faint components that biased lower-resolution observations. The distribution of fractional widths is similar to that of rings in protoplanetary discs, but the wide population is still more prevalent than in protoplanetary discs as found in REASONS \citep{Matra2025}.

\subsection{ARKS III: The vertical structure of debris discs, \cite{ver_arks}}

In this paper, we analyse the vertical distribution of dust using parametric models. We resolve the vertical structure of 13 belts, finding a wide distribution of vertical aspect ratios ($H/r$), with values ranging between 0.003--0.2, with a median of 0.02. Moreover, we find that for most belts, their vertical distribution approximates better to a Lorentzian than a Gaussian distribution, hinting at the presence of multiple dynamical populations. The inferred dynamical excitation of these belts could be explained by self-stirring, with half of them requiring belt masses below 20 $M_{\oplus}$. 

\subsection{ARKS IV: CO gas imaging and overview, \cite{gas_arks}}

In this paper, we present and analyse the $^{12}$CO and $^{13}$CO J=3-2 CO gas observations of HD~9672, HD~32297, HD~121617, HD~131488, and HD~131835. We describe how the image cubes were generated and how their integrated and peak intensity maps were calculated. We analyse the spatial distribution of CO and compare it with that of the millimetre dust, finding that the CO gas intensity is broader than the dust emission and its peak is shifted inwards in comparison to the dust, though the offset between the dust and the gas peaks varies with system. If the gas is secondary, this could be evidence of viscous spreading. However, these differences could at least be partially explained by the high optical depth of CO emission.

We present radially resolved $^{12}$CO/$^{13}$CO isotopologue ratios for the five systems with new ARKS observations (i.e. excluding HD~39060/$\beta$~Pic) and find that the $^{12}$CO and $^{13}$CO in all but one system, HD~9672 (49~Ceti), must both be either optically thin or optically thick. Using spectrospatial stacking, we measured the integrated line flux of $^{12}$CO and $^{13}$CO in each gas-rich system and conducted a deep search for $^{12}$CO in every disc that did not previously have a CO detection. No new CO detections were made, although this allowed us to constrain upper limits on the $^{12}$CO integrated line flux in all systems without detected CO.

\subsection{ARKS V: Comparison between scattered light and thermal emission, \cite{scat_arks}}

In addition to ARKS observations, we have collected scattered light data of all systems in our sample, tracing the distribution of small micron-sized grains. Of the 24 belts, 15 have been detected in scattered light. The reported detections were made with at least one of these three facilities: the Very Large Telescope (VLT) SPHERE instrument, the Gemini Planet Imager (GPI), or the Hubble Space Telescope (HST), noting that the entire ARKS sample has been observed by the VLT/SPHERE instrument. In this paper, we focus on those 15 detections and present the data reduction, forward-modelling, and comparison of the spatial distribution of small and large grains. For gas-poor systems, we find only subtle differences in peak surface density locations. However, for gas-rich systems, we find that the distribution of small grains is significantly shifted outwards by ${\gtrsim}10\%$, suggesting that the interplay of radiation pressure and gas drag has a significant effect on the dynamics of $\mu$m-grains \citep[][]{Takeuchi2001, Krivov2009}. We also detect for the first time the scattered light emission of the belt surrounding the K-type star TYC~9340-437-1 with HST/NICMOS. Lastly, we summarise the mass and orbital properties of the known planets in the 24 systems, their location relative to the disc edges and gaps, and present additional constraints from Gaia astrometry, Gaia-Hipparcos proper motion anomalies, and direct imaging.

\subsection{ARKS VI: Asymmetries and offsets, \cite{asym_arks}}

In this paper, we investigate whether the (sub)millimetre dust emission of the sample contains asymmetries, and if the discs contain stellocentric offsets (eccentricities). We use a series of empirical diagnostics to search for different types of asymmetric features in the data. We find that 10/24 systems present significant asymmetries in the form of disc eccentricities, arcs, and global emission asymmetries (i.e. along either their major or minor axes, or azimuthally). Tentative asymmetries (at the $3{-}5\sigma$ level) are found in four other discs. 
We characterise these asymmetries and briefly discuss plausible dynamical scenarios that could explain these features.
We find that the presence of an asymmetry/offset in the ARKS sample appears to be correlated with the fractional luminosity of cold dust.
We also find a tentative enhancement in the fraction of systems hosting a continuum asymmetry, and those that are CO-rich. 
Overall, this study, the first (sub)millimetre population analysis of debris disc asymmetries, highlights that asymmetries and offsets in debris discs are likely common.

\subsection{ARKS VII: Optically thick gas in the HD~121617 disc with broad CO Gaussian local line profiles, \cite{line_arks}}

In this paper, we analyse the intrinsic line profile of $^{12}$CO in the HD~121617 disc. The observed line profile is very broad, displaying a FWHM of 1.1~km/s, which could be interpreted as the gas having a very high kinetic temperature of ${\sim}380$~K at 73~au. However, the integrated intensity images (moment 0's) of both $^{12}$ and $^{13}$CO display an azimuthal modulation that is typical of optically thick discs (also seen in the other 2 non-edge-on gas-rich discs in ARKS). Using radiative transfer models, we show how a massive (${\gtrsim}0.1~M_{\oplus}$), cold (${\sim}38$~K), and narrow (FWHM of 17~au) ring of CO gas can reproduce the line width, moment 0 and peak intensity images. These results confirm the previous findings of a low excitation temperature for CO gas \citep[e.g.][]{Cataldi2023} and that the $^{12}$CO emission must be optically thick. Additionally, these results would imply that CO is thermally decoupled from the dust, displaying a significantly lower temperature.

\subsection{ARKS VIII: A dust arc and non-Keplerian gas kinematics in HD 121617, \cite{hd121617_arks}}

In this paper, we investigate the asymmetric dust distribution and CO kinematics in the HD~121617 disc. We find that the dust arc has a morphology similar to that attributed to vortices in protoplanetary discs, and that it is absent or much less pronounced in the distribution of small grains and gas (although the CO gas emission is optically thick and thus could hide an arc in the gas distribution). The CO gas kinematics show strong deviations from Keplerian rotation due to strong pressure gradients at the inner and outer edges of the ring. We retrieve the rotation curve and use it to derive profiles for the pressure gradient and gas surface density. We can reconcile this surface density with that of the CO gas depending on the assumed stellar mass and gas sound speed (determined by the gas temperatures and mean molecular weight). If the gas densities are high enough, requiring a primordial origin, the dust radial confinement and azimuthal arc may result from dust grains responding to gas drag \citep{vortex_arks}. Alternatively, the asymmetry may be due to planet-disc interactions via mean motion resonances (Pearce et al. in prep).

\subsection{Paper IX: Gas-driven origin for continuum arc in the debris disc HD 121617, \cite{vortex_arks}}

A key finding presented in \cite{asym_arks} and \cite{hd121617_arks} is that the mm-emission from the dust ring around HD~121617 shows a significant asymmetry with an arc of increased emission. This suggests that dust accumulates at a preferred azimuth within the ring, which itself is embedded in a gas ring with steep pressure gradients on either side. We explore two explanations for this feature: In \citet{vortex_arks}, we investigate whether a shallow gas vortex, similar to those seen in protoplanetary discs, could generate such an azimuthal dust over-density. Through hydrodynamical simulations with varying gas masses, we find that this scenario only works if the total gas mass in HD~121617 is around ${\sim}25\,M_\oplus$, roughly 10$^2$ times the estimated minimum CO mass. Hence, this scenario implies a hydrogen-dominated composition and a primordial origin for the CO. By contrast, an alternative explanation involving an outward-migrating planet can also account for the observed asymmetry, as it would trap dust grains in mean-motion resonances; this alternative will be explored in a forthcoming study (Pearce et al., in prep.).

\subsection{Paper X: Interpreting the peculiar dust rings around HD 131835, \cite{hd131835_arks}}

HD~131835 contains at least two distinct rings, with the outermost being brightest in scattered light, indicating that micron-sized grains reside mostly in that outer ring, while the innermost is much brighter at millimetre wavelengths, indicating that millimetre-sized grains instead reside mostly in the inner ring \citep{scat_arks}.
In this paper, we explore two possible explanations for this grain size segregation. We
(i) use collisional models to test under which conditions two planetesimal belts can produce two dust rings with such different properties, and (ii) use dynamical models of dust migration under the influence of gas drag
to investigate if the gas present in this disc could explain this behaviour instead. We find that gas drag can form an outer ring out of small dust \citep[as in][]{Takeuchi2001}, but the simple dynamical model cannot reproduce the brightness of HD~131835's outermost ring. Nevertheless, the gas-driven explanation is promising, and we discuss how a more comprehensive model may change this result. The collisional scenario might reproduce observations, although it requires an extreme difference in the dynamical excitation and/or material strength between the two rings, which remains to be explained.

\section{Conclusions}
\label{sec:conclusions}

The ARKS ALMA large programme was motivated by a series of questions that have arisen over the last decade while studying exoKuiper belts. Below, we summarise these questions, how ARKS has contributed to answering them, and what new questions have arisen.

\paragraph{What type of radial substructures are present in exoKuiper belts?} ARKS has shown that there is a great diversity of radial structures. Up to one-third of the belts in ARKS show substructures in the form of multiple rings or local maxima with gaps of varied depths in between, one-third display narrow rings surrounded by low-amplitude additional rings or halos, and the rest are consistent with wide and smooth single belts \citep{rad_arks}. The high fraction of multi-ring belts and narrow single rings (${\sim}60$\%) suggests that a high fraction of exoKuiper belts may inherit their solid distribution from that in protoplanetary discs that predominantly show substructures in the form of multiple or single narrow rings. Alternatively, these structures may appear at a later stage (e.g. gaps cleared by planets). The other ${\sim}40$\% of wide and smooth belts may have formed or evolved in very different ways, for example in migrating protoplanetary rings or scattered by planets. Finally, ARKS has also shown a diversity in the steepness of belt edges. Belt inner edges within 100~au tend to be steep and consistent with planet sculpting, whereas inner edges at larger distances tend to be shallower and consistent with being shaped by collisional evolution. We do not find clear correlations between these structures and the system properties (e.g. age, spectral type, presence of planets), which leaves the origin of these structures unconstrained.

\paragraph{How dynamically excited are exoKuiper belts?} ARKS has shown that belts have a wide range of dynamical excitation as inferred from their vertical thickness \citep{ver_arks}. These levels overlap those measured for the Kuiper belt, from a few degrees to tens of degrees. Moreover, ARKS has shown that the dust vertical distributions are better reproduced by non-Gaussian distributions such as a Lorentzian, which suggests the presence of multiple dynamical populations. Similar to the radial structures, we do not find clear correlations between the derived vertical structures and the system properties. However, we do find that the radial and vertical widths of belts are correlated, which may indicate that the processes that excite the orbits of solids are also responsible for making these belts wider. Despite these important insights, it is still an open question whether the dynamical excitation is set by planets not embedded in the belts or by dwarf planets within the belts.

\paragraph{Are asymmetries common in exoKuiper belts?} ARKS has shown that asymmetries in the dust distribution are common;  10  of the 24 belts show a significant asymmetry in the form of arcs, belt eccentricities, and warps \citep{asym_arks}. Whether these asymmetries are caused by planet--disc interactions, stellar flybys, or gas drag is an open question. Asymmetries in the gas distribution are yet to be examined in detail, but at least one system shows tentative evidence of an eccentricity in its gas distribution \citep[HD~121617,][]{hd121617_arks}, while other edge-on ones show azimuthal asymmetries \citep[HD~39060/$\beta$~Pic, HD32297, HD131488,][]{gas_arks}.

\paragraph{What is the origin of the gas?} ARKS has shown that the CO gas emission spans a larger range of radii than the dust, with a peak intensity slightly closer to the star than the dust \citep{gas_arks}. This may be partly explained by the CO emission being optically thick, as demonstrated for one system by \cite{line_arks}. However, in a few systems, CO emission is detected close to the star, which strongly suggests that the gas and dust distributions are different. This wider span is a feature seen in primordial and secondary gas models. In the former, the dust distribution can be shaped by gas drag, producing a narrower distribution of large grains trapped in pressure maxima if gas densities are high enough \citep{vortex_arks}. In the secondary scenario, CO gas is expected to have a wider distribution if it is shielded, allowing it to viscously spread before being photodissociated \citep{Kral2019, Marino2020gas}.

\paragraph{Does gas affect the dust dynamics?} ARKS has found that in gas-bearing systems, the spatial distribution of micron-sized grains is significantly shifted outwards compared to millimetre-sized grains \citep{scat_arks}. This is a feature expected in optically thin gas-rich discs, due to the combined effect of radiation pressure and gas drag \citep{hd131835_arks}. Moreover, for one of these gas-bearing belts, we found an overdensity in the distribution of millimetre-sized dust that resembles the expected morphology of dust trapped in a vortex \citep{asym_arks, hd121617_arks, vortex_arks}. These findings suggest that gas may play an important role in shaping the distribution of small and large dust in gas-bearing exoKuiper belts. These results have also triggered the question of whether the wider span of gas relative to dust is a consequence of gas viscous spreading or dust trapping. 

\paragraph{Does the gas display non-Keplerian kinematics?} For at least one system, HD~121617, we find strong deviations from Keplerian rotation due to strong pressure gradients that are consistent with the CO gas intensity distribution \citep{hd121617_arks}. These kinematic features, combined with the intensity distribution and radiative transfer models, could allow us to break degeneracies and determine the mean molecular weight of the gas. 

Perhaps the most important question that is yet to be answered is whether any or most of the observed structures are linked to the presence of planets in these systems. Some of these systems host planets, but most reside far from the edges of these belts. The James Webb Space Telescope and soon the Extremely Large Telescope may reveal or rule out the presence of planets actively shaping these discs. 

\section*{Data availability}
The ARKS data used in this paper and others can be found in the \href{https://dataverse.harvard.edu/dataverse/arkslp}{ARKS dataverse}. The continuum data products produced by this work can be found within the \href{https://dataverse.harvard.edu/dataset.xhtml?persistentId=doi:10.7910/DVN/VNGHPQ&}{ARKS I's dataset} (\href{https://doi.org/10.7910/DVN/VNGHPQ}{doi.org/10.7910/DVN/VNGHPQ}). The data reduction scripts can be found at \href{https://github.com/SebaMarino/ARKS-data-reduction}{github.com/SebaMarino/ARKS-data-reduction} (\href{https://doi.org/10.5281/zenodo.17432148}{doi.org/10.5281/zenodo.17432148}). For more information, visit \href{https://arkslp.org}{arkslp.org}.

\begin{acknowledgements}

We thank the referee for their constructive review of this manuscript that helped improve the quality and clarity of this work. This paper makes use of the following ALMA data: ADS/JAO.ALMA\# 2022.1.00338.L, 2012.1.00142.S, 2012.1.00198.S, 2015.1.01260.S, 2016.1.00104.S, 2016.1.00195.S, 2016.1.00907.S, 2017.1.00167.S, 2017.1.00825.S, 2018.1.01222.S and 2019.1.00189.S. ALMA is a partnership of ESO (representing its member states), NSF (USA) and NINS (Japan), together with NRC (Canada), MOST and ASIAA (Taiwan), and KASI (Republic of Korea), in cooperation with the Republic of Chile. The Joint ALMA Observatory is operated by ESO, AUI/NRAO and NAOJ. The National Radio Astronomy Observatory is a facility of the National Science Foundation operated under cooperative agreement by Associated Universities, Inc. The project leading to this publication has received support from ORP, that is funded by the European Union’s Horizon 2020 research and innovation programme under grant agreement No 101004719 [ORP]. We are grateful for the help of the UK node of the European ARC in answering our questions and producing calibrated measurement sets. This research used the Canadian Advanced Network For Astronomy Research (CANFAR) operated in partnership by the Canadian Astronomy Data Centre and The Digital Research Alliance of Canada with support from the National Research Council of Canada the Canadian Space Agency, CANARIE and the Canadian Foundation for Innovation.
SM acknowledges funding by the Royal Society through a Royal Society University Research Fellowship (URF-R1-221669) and the European Union through the FEED ERC project (grant number 101162711). LM acknowledges funding by the European Union through the E-BEANS ERC project (grant number 100117693), and by the Irish research Council (IRC) under grant number IRCLA- 2022-3788. Views and opinions expressed are however those of the author(s) only and do not necessarily reflect those of the European Union or the European Research Council Executive Agency. Neither the European Union nor the granting authority can be held responsible for them. AMH acknowledges support from the National Science Foundation under Grant No. AST-2307920. CdB acknowledges support from the Spanish Ministerio de Ciencia, Innovaci\'on y Universidades (MICIU) and the European Regional Development Fund (ERDF) under reference PID2023-153342NB-I00/10.13039/501100011033, from the Beatriz Galindo Senior Fellowship BG22/00166 funded by the MICIU, and the support from the Universidad de La Laguna (ULL) and the Consejer\'ia de Econom\'ia, Conocimiento y Empleo of the Gobierno de Canarias. AB acknowledges research support by the Irish Research Council under grant GOIPG/2022/1895. MRJ acknowledges support from the European Union's Horizon Europe Programme under the Marie Sklodowska-Curie grant agreement no. 101064124 and funding provided by the Institute of Physics Belgrade, through the grant by the Ministry of Science, Technological Development, and Innovations of the Republic of Serbia. JBL acknowledges the Smithsonian Institute for funding via a Submillimeter Array (SMA) Fellowship, and the North American ALMA Science Center (NAASC) for funding via an ALMA Ambassadorship. SMM acknowledges funding by the European Union through the E-BEANS ERC project (grant number 100117693), and by the Irish research Council (IRC) under grant number IRCLA- 2022-3788. Views and opinions expressed are however those of the author(s) only and do not necessarily reflect those of the European Union or the European Research Council Executive Agency. Neither the European Union nor the granting authority can be held responsible for them. JM acknowledges funding from the Agence Nationale de la Recherche through the DDISK project (grant No. ANR-21-CE31-0015) and from the PNP (French National Planetology Program) through the EPOPEE project. PW acknowledges support from FONDECYT grant 3220399 and ANID -- Millennium Science Initiative Program -- Center Code NCN2024\_001. Support for BZ was provided by The Brinson Foundation. RBW was supported by a Royal Society Grant (RF-ERE-221025). This material is based upon work supported by the National Science Foundation Graduate Research Fellowship under Grant No. DGE 2140743. EM acknowledges support from the NASA CT Space Grant. TDP is supported by a UKRI Stephen Hawking Fellowship and a Warwick Prize Fellowship, the latter made possible by a generous philanthropic donation. This work was also supported by the NKFIH research grant K-147380. MB acknowledges funding from the Agence Nationale de la Recherche through the DDISK project (grant No. ANR-21-CE31-0015). EC acknowledges support from NASA STScI grant HST-AR-16608.001-A and the Simons Foundation. S.E. is supported by the National Aeronautics and Space Administration through the Exoplanet Research Program (Grant No. 80NSSC23K0288, PI: Faramaz). This work was also supported by the NKFIH NKKP grant ADVANCED 149943 and the NKFIH excellence grant TKP2021-NKTA-64. Project no.149943 has been implemented with the support provided by the Ministry of Culture and Innovation of Hungary from the National Research, Development and Innovation Fund, financed under the NKKP ADVANCED funding scheme. JPM acknowledges research support by the National Science and Technology Council of Taiwan under grant NSTC 112-2112-M-001-032-MY3. SP acknowledges support from FONDECYT Regular 1231663 and ANID -- Millennium Science Initiative Program -- Center Code NCN2024\_001. A.A.S. is supported by the Heising-Simons Foundation through a 51 Pegasi b Fellowship.

\end{acknowledgements}

\bibliographystyle{aa}
\bibliography{bib}

\appendix

\section{Observation details}

\begin{table*}

    \caption{All new and archival observations used for ARKS. \label{tab:obs}}
    \centering
    \begin{adjustbox}{max width=0.98\textwidth}
    \begin{threeparttable}
    \begin{tabular}{ l c c c c c c c c c c c }
    \hline
    \hline 
    Name  & $\Delta R$  & $\Delta z\sin(i)$  & Beam & rms & Band & Antenna & $N_{\rm exe}$ & ToS & Date range & Archival   \\
        & [au ($\arcsec$)] & [au ($\arcsec$)]  & [au ($\arcsec$)] & [$\mu$Jy/beam] &  &  conf.  &             &  [h]  &         &  data  \\
    \hline
HD 9672$^{\rm g}$ & 147(2.6) & 16(0.27) & 9(0.16) & 12 & 7 & 5-6,2 & 6,1  & 3.1, 0.8 & 10/2018-06/2024 & 1 \\
HD 10647 & 70(4.0) & 11(0.66) & 8(0.43) & 10 & 7 & 4,1,7m & 6,3,6 & 4.1, 2.0, 4.4  & 07/2016-05/2024 & 2,3 \\
HD 14055 & 160(4.5) & 21(0.59) & 8(0.23) & 10 & 7 & 4,1,7m & 11,0,21  & 8.5,0,17.3 & 10/2022-07/2024 &  \\
HD 15115 & 21(0.4) & 11(0.22) & 6(0.12) & 6 & 7 & 6,3 & 12,3  & 9.5,1.8 & 10/2022-06/2024 &  \\
HD 15257 & 220(4.5) & 20(0.42) & 36(0.74) & 24 & 7 & 2,7m & 4,11  & 2.7,8.8 & 10/2022 &  \\
HD 32297$^{\rm g}$ & 62(0.5) & 14(0.11) & 7(0.06) & 17 & 7 & 7,4 & 4,2  & 2.6,1.0 & 06/2023-12/2023 &  \\
\rowcolor{gray!10} HD 39060$^{\rm g}$ & 92(4.7) & 12(0.63) & 4(0.22) & 13 & 6 & 5,2,7m & 3,1,2 & 1.9,0.5,0.9 & 10/2013-08/2015 & 4 \\
HD 61005 & 38(1.0) & 9(0.23) & 6(0.15) & 21 & 7 & 6,3 & 3,1  & 1.7,0.3 & 10/2022-05/2023 &  \\
HD 76582 & 210(4.3) & 25(0.50) & 27(0.56) & 18 & 7 & 2,7m & 2,9  & 1.7,7.0 & 10/2022-12/2022 &  \\
HD 84870 & 260(2.9) & 23(0.26) & 49(0.56) & 14 & 7 & 3,7m & 7,14  & 5.8,10.5 & 10/2022-12/2022 &  \\
\rowcolor{gray!10} HD 92945 & 80(3.7) & 10(0.48) & 9(0.40) & 17 & 7 & 3,7m & 3,5  & 2.4, 4.1 & 10/2016-03/2017 & 5 \\
HD 95086 & 180(2.1) & 12(0.14) & 30(0.34) & 19 & 7 & 3 & 3  & 2.4 & 01/2023 &  \\
\rowcolor{gray!10} HD 107146 & 110(4.0) & 5(0.17) & 16(0.59) & 13 & 7 & 5,3,2,7m & 1,1,2,4 & 0.7,0.8,1.6,2.3 & 10/2016-05/2021 & 6,7 \\
\rowcolor{gray!10} HD 107146 & 110(4.0) & 5(0.17) & 18(0.65) & 6 & 6 & 3 & 5 & 4.0 & 04/2017 & 8 \\
HD 109573 & 15(0.2) & 9(0.13) & 6(0.08) & 20 & 7 & 7,4 & 2,1 & 1.0,0.2 & 01/2023-06/2023 &  \\
HD 121617$^{\rm g}$ & 60(0.5) & 6(0.05) & 14(0.12) & 11 & 7 & 6,3 & 7,2  & 5.8,1.4 & 10/2022-05/2023 &  \\
HD 131488$^{\rm g}$ & 46(0.3) & 11(0.07) & 6(0.04) & 9 & 7 & 8,5 & 6,3  & 4.8,1.9 & 04/2023-06/2023 &  \\
HD 131835$^{\rm g}$ & 87(0.7) & 9(0.07) & 17(0.13) & 11 & 7 & 6,3 & 9,2  & 6.6,1.6 & 12/2022-05/2023 &  \\
HD 145560 & 50(0.4) & 7(0.05) & 11(0.09) & 9 & 7 & 6,3 & 7,3  & 5.5,2.0 & 10/2022-05/2023 &  \\
HD 161868 & 110(3.7) & 14(0.46) & 15(0.52) & 19 & 7 & 2,7m & 3,11  & 2.0,7.9 & 10/2022 &  \\
HD 170773 & 68(1.8) & 12(0.34) & 12(0.33) & 22 & 7 & 3,7m & 2,3  & 1.0,2.5 & 10/2022-01/2023 &  \\
\rowcolor{gray!10} HD 197481 & 12(1.3) & 4(0.42) & 3(0.35) & 16 & 6 & 5,2 & 2,1  & 0.9,0.5 & 03/2014-06/2015 & 9 \\
\rowcolor{gray!10} HD 206893 & 100(2.5) & 5(0.13) & 18(0.45) & 6 & 6 & 5,2 & 4,2 & 3.2,1.3 & 06/2018-09/2018 & 10 \\
TYC 9340-437-1 & 100(2.7) & 10(0.27) & 20(0.55) & 24 & 7 & 3,7m & 2,4  & 1.7,3.0 & 10/2022 &  \\
\rowcolor{gray!10} HD 218396 & 250(6.1) & 21(0.53) & 35(0.85) & 11 & 7 & 1,7m & 6,30  & 4.6,21.7 & 10/2016-06/2018 & 11 \\
    \hline

    \end{tabular}

    \tablefoot{ Column 2 shows the estimated radial FWHM from REASONS. Column 3 shows the expected projected vertical FWHM, where we assume $h=0.05$. Column 4 shows the clean beam size obtained after imaging all the data using a robust value of 0.5 (see Sect.~\ref{sec:imaging}) and averaging the beam major and minor axes. Column 5 shows the corresponding image rms for the robust=0.5 images. The last six columns show the band, the approximate antenna configurations, the number of executions per configuration, the time on source per configuration, the range of observing dates, and the references to archival data that we use, respectively. Systems with archival data only and not observed by ARKS are highlighted in grey, and systems with CO gas are flagged `$^{\rm g}$'. \\
    \textbf{Archival data used}: (1): 2018.1.01222.S \citep{Cataldi2023, Delamer2023}; (2,3): 2015.1.01260.S, 2017.1.00167.S \citep{Lovell2021}; (4): 2012.1.00142.S \citep{Matra2019}; (5): 2016.1.00104.S \citep{Marino2019}; (6,7): 2016.1.00104.S, 2019.1.00189.S \citep{Marino2018, Imaz-Blanco2023}; (8): 2016.1.00195.S \citep{Marino2018}; (9): 2012.1.00198.S \citep{Daley2019}; (10): 2017.1.00825.S \citep{Nederlander2021, Marino2020hd206}; (11): 2016.1.00907.S \citep{Faramaz2021}.
    }
    \end{threeparttable}
    \end{adjustbox}

\end{table*}

Table~\ref{tab:obs} presents a summary of all the new and archival observations used for ARKS.

\section{Data correction parameters}

Table~\ref{tab:mcmc_results} presents the best-fit disc parameters for the model used to correct the data as described in Sect.~\ref{sec:correction}.

\begin{table*}

    \caption{Continuum best-fit parameters \label{tab:mcmc_results}} 
    
    \centering
    \begin{adjustbox}{max width=0.98\textwidth}
    \begin{threeparttable}

    \begin{tabular}{l c c c c c c c c c }
    \hline
    \hline 
    Name  & $R$  & $\Delta R$  & $h$ & $i$     & PA      & $M_{\rm dust}$  & $F_{\star}$ & $\Delta \mathrm{RA}$ & $\Delta \mathrm{Dec}$ \\
          & [au]        & [au] &     & [\degr] & [\degr] & [$M_{\oplus}$]  & [$\mu$Jy] & [mas] & [mas] \\
    \hline
HD 9672$^{\rm g}$ & $135.2\pm1.9$ & $158\pm3$ & $0.066\pm0.004$ & $78.7\pm0.2$ & $107.9\pm0.2$ & $0.21$ & $<19$ (28) & $68\pm76$ & $-33\pm30$ \\
HD 10647 & $103.6\pm0.4$ & $65\pm2$ & $0.062\pm0.003$ & $77.8\pm0.1$ & $57.3\pm0.1$ & $0.03$ & $131\pm9$ (96) & $90\pm56$ & $28\pm51$ \\
HD 14055 & $169.4\pm2.9$ & $162\pm9$ & $0.048\pm0.007$ & $80.7\pm0.2$ & $162.2\pm0.3$ & $0.05$ & $87\pm11$ (111) & $-57\pm59$ & $-84\pm155$ \\
HD 15115 & $80.7\pm0.3$ & $14\pm1$ & $0.017\pm0.001$ & $86.7\pm0.0$ & $98.5\pm0.0$ & $0.07$ & $24\pm4$ (22) & $44\pm12$ & $-8\pm5$ \\
HD 15257 & $184.2\pm20.2$ & $261\pm41$ &  & $59.4\pm4.9$ & $47.3\pm5.5$ & $0.06$ & $79\pm35^{\rm m}$ (73) & $86\pm134$ & $14\pm180$ \\
HD 32297$^{\rm g}$ & $115.9\pm0.3$ & $37\pm1$ & $0.011\pm0.001$ & $88.3\pm0.0$ & $47.5\pm0.0$ & $0.87$ & $24\pm13^{\rm m}$ (3) & $19\pm4$ & $18\pm4$ \\
HD 39060$^{\rm g}$ & $104.3\pm1.4$ & $97\pm3$ & $0.059\pm0.003$ & $86.4\pm0.2$ & $29.9\pm0.1$ & $0.12$ & $86\pm14$ (70) & $-54\pm14$ & $-2\pm20$ \\
HD 61005 & $72.4\pm0.2$ & $36\pm1$ & $0.020\pm0.002$ & $85.9\pm0.1$ & $70.4\pm0.1$ & $0.15$ &  (12) & $8\pm22$ & $-4\pm8$ \\
HD 76582 & $203.6\pm4.9$ & $176\pm9$ &  & $72.6\pm0.7$ & $104.7\pm0.7$ & $0.12$ &  (37) & $100\pm46$ & $54\pm33$ \\
HD 84870 & $199.3\pm7.8$ & $232\pm15$ &  & $47.1\pm2.6$ & $-25.7\pm4.1$ & $0.13$ &  (9) & $81\pm57$ & $-103\pm63$ \\
HD 92945 & $78.2\pm1.0$ & $73\pm6$ & $0.032\pm0.013^{\rm m}$ & $65.4\pm0.6$ & $100.0\pm0.6$ & $0.06$ & $45\pm13$ (25) & $-84\pm55$ & $12\pm35$ \\
HD 95086 & $197.0\pm4.6$ & $173\pm8$ &  & $31.6\pm3.5$ & $100.0\pm5.7$ & $0.49$ &  (9) & $-91\pm45$ & $127\pm34$ \\
HD 107146 & $84.2\pm0.2$ & $89\pm1$ &  & $19.3\pm0.6$ & $153.2\pm1.7$ & $0.22$ &  (29) & $36\pm35$ & $-48\pm35$ \\
HD 109573 & $75.9\pm0.1$ & $7\pm0$ & $<0.02$ & $76.6\pm0.1$ & $26.5\pm0.0$ & $0.23$ &  (20) & $11\pm2$ & $-32\pm2$ \\
HD 121617$^{\rm g}$ & $76.3\pm0.4$ & $18\pm1$ &  & $44.1\pm0.6$ & $58.7\pm0.7$ & $0.21$ &  (6) & $8\pm8$ & $14\pm7$ \\
HD 131488$^{\rm g}$ & $89.7\pm0.1$ & $13\pm1$ & $0.011\pm0.001$ & $85.0\pm0.1$ & $97.2\pm0.0$ & $0.58$ &  (3) & $0\pm2$ & $1\pm1$ \\
HD 131835$^{\rm g}$ & $77.1\pm0.5$ & $65\pm2$ & $0.049\pm0.005$ & $74.2\pm0.2$ & $59.2\pm0.2$ & $0.45$ &  (5) & $6\pm6$ & $3\pm6$ \\
HD 145560 & $75.4\pm0.5$ & $29\pm2$ & $<0.11$ & $47.6\pm0.6$ & $39.4\pm0.9$ & $0.28$ &  (4) & $15\pm12$ & $-9\pm11$ \\
HD 161868 & $121.2\pm3.7$ & $134\pm8$ &  & $66.1\pm1.5$ & $57.6\pm1.6$ & $0.03$ & $161\pm18$ (146) & $49\pm40$ & $-131\pm42$ \\
HD 170773 & $191.9\pm2.6$ & $66\pm4$ &  & $32.9\pm1.9$ & $113.0\pm2.9$ & $0.16$ & $84\pm21$ (39) & $-77\pm40$ & $26\pm35$ \\
HD 197481 & $33.6\pm0.2$ & $13\pm1$ & $0.015\pm0.003$ & $88.3\pm0.1$ & $128.6\pm0.1$ & $0.012$ & $245\pm14$ (50) & $32\pm15$ & $6\pm13$ \\
HD 206893 & $105.1\pm6.0$ & $116\pm10$ &  & $45.2\pm2.5$ & $60.3\pm3.6$ & $0.04$ &  (12) & $40\pm132$ & $53\pm106$ \\
TYC 9340-437-1 & $95.7\pm2.5$ & $99\pm5$ &  & $18.0\pm8.0$ & $149.2\pm20.2$ & $0.17$ &  (10) & $81\pm65$ & $-98\pm64$ \\
HD 218396 & $240.4\pm4.6$ & $182\pm9$ &  & $28.8\pm3.3$ & $49.7\pm5.7$ & $0.11$ & $59\pm10$ (35) & $-41\pm65$ & $-57\pm69$ \\
    \hline 
    \end{tabular}

    \tablefoot{Best-fit parameters derived from the MCMC sampling, using the median and the 16th and 84th percentiles to estimate the uncertainties. For sources fitted with a double Gaussian (HD~107146, HD~92945, and HD~15115), $R$ represents the average between the two Gaussian centres and $\Delta R$ is the sum of the FWHMs of the two Gaussians. Column 8 shows the stellar flux at the shortest available ALMA wavelength as derived from the MCMC if it was included as a free parameter and the photospheric flux expected from extrapolating the stellar SED in brackets. For marginalised distributions of $h$ and $F_{\star}$ peaking at zero, we present upper limits estimated with the 99.7 percentile. If their marginalised distributions peaked above zero but still extended to zero, we flagged those marginal values  `$^{\rm m}$'. The last two columns show the best-fit right ascension and declination offsets for the long baseline execution used as a reference. Systems with CO gas are flagged `$^{\rm g}$'. }

    \end{threeparttable}
    \end{adjustbox}
    
\end{table*}

\section{Fundamental stellar parameters}
\label{sec:star}

In order to determine the fundamental stellar parameters for the ARKS sample of stars, we applied the Bayesian inference code of \citet{delburgo2016,delburgo2018} 
on a grid of PARSEC 1.2S stellar evolution models \citep{bressan2012,Chen2014, Chen2015,Tang2014}. This choice is based on the reasonably good statistical match of the measured dynamical masses with the corresponding predictions for detached eclipsing binaries. In particular, for binary stars on the main-sequence, with a small discrepancy of\,4\% on average \citep[][]{delburgo2018}. 

The grid of PARSEC v1.2S models arranged for this analysis comprises ages ranging from 2 to 13,800 million years and steps of 5\%, and [M/H] from -2.18 to 0.51 with steps of 0.02 dex, adopting the photometric passband calibration of \citet{riello2021}, 
with the zero points of the VEGAMAG system. We used this grid to infer the stellar fundamental parameters of all stars through the aforementioned Bayesian code, fed by the following three input parameters: the absolute $G$ magnitude $M_{G}$, the colour $G_{\rm BP}-G_{\rm RP}$, and the stellar age compiled from the literature. We did not constrain the stellar evolution stage. Given the relatively small distances, we assumed null extinction when deriving the colour $G_{\rm BP}-G_{\rm RP}$ and $M_{G}$ from the Gaia DR3 photometry and astrometry \citep[][]{gaiadr3}. $M_{G}$ was determined from the apparent G magnitude by subtracting the distance modulus, which was calculated from the Gaia DR3 trigonometric parallaxes. The resulting stellar parameters are shown in Table~\ref{tab:stellarparam}.

\begin{table*}
    \caption{Fundamental stellar parameters of the 24 host stars of the ARKS sample.}
    \centering
    \begin{tabular}{l c c c c c c c c}
    \hline
    \hline
      Name  & $T_{\rm eff}$ & $M_{\star}$   & $R_{\star}$ & $L_{\star}$  &  $\log(g)$  & [Fe/H] \\
            &  [K]  & [$M_\odot$] & [$R_\odot$] &  [$L_\odot$] & [cgs] & [dex] \\
     \hline
HD 9672       & 8924$\pm$32      &  1.9955$\pm$0.0004       &  1.643$\pm$0.007         &  15.43$\pm$0.11          &  4.307$\pm$0.003         &    0.182$\pm$0.013 \\
HD 10647      & 6154$\pm$15      &  1.116$\pm$0.019         &  1.085$\pm$0.005         &  1.521$\pm$0.006         &  4.415$\pm$0.010         &    0.02$\pm$0.05 \\
HD 14055      & 9120$\pm$64      &  2.19$\pm$0.09           &  2.06$\pm$0.04           &  26.3$\pm$0.9            &  4.153$\pm$0.026         &    0.21$\pm$0.19 \\
HD 15115      & 6764$\pm$15      &  1.426$\pm$0.005         &  1.380$\pm$0.005         &  3.592$\pm$0.016         &  4.312$\pm$0.004         &    0.19$\pm$0.03 \\
HD 15257      & 7256$\pm$35      &  1.75$\pm$0.13           &  2.37$\pm$0.04           &  13.99$\pm$0.28          &  3.93$\pm$0.05           &    0.1$\pm$0.3  \\
HD 32297      & 7874$\pm$22      &  1.57$\pm$0.06           &  1.451$\pm$0.013         &  7.29$\pm$0.10           &  4.311$\pm$0.024         &   -0.15$\pm$0.24 \\
HD 39060   & 7951$\pm$32      &  1.724$\pm$0.005         &  1.534$\pm$0.008         &  8.47$\pm$0.06           &  4.303$\pm$0.005         &    0.157$\pm$0.018 \\
HD 61005      & 5577$\pm$17      &  0.95$\pm$0.007          &  0.834$\pm$0.006         &  0.6060$\pm$0.0011        &  4.573$\pm$0.008         &    0.01$\pm$0.03 \\
HD 76582      & 7796$\pm$21      &  1.61$\pm$0.15           &  1.715$\pm$0.02          &  9.79$\pm$0.17           &  4.17$\pm$0.05           &   -0.1$\pm$0.4  \\
HD 84870      & 7741$\pm$20      &  1.66$\pm$0.04           &  1.566$\pm$0.011         &  7.93$\pm$0.06           &  4.268$\pm$0.016         &    0.11$\pm$0.13 \\
HD 92945      & 5175$\pm$9       &  0.850$\pm$0.007         &  0.747$\pm$0.004         &  0.3609$\pm$0.0013       &  4.62$\pm$0.008          &   -0.00$\pm$0.06 \\ 
HD 95086      & 7674$\pm$22      &  1.541$^{+0.020}_{-0.002}$  &  1.455$^{+0.007}_{-0.005}$ &  6.614$^{+0.019}_{-0.014}$ &  4.3$^{+0.008}_{-0.003}$   &   -0.19$^{+0.07}_{-0.02}$ \\
HD 107146     & 5885$^{+22}_{-15}$ &  1.04$^{+0.03}_{-0.01}$   &  0.935$^{+0.012}_{-0.006}$ &  0.944$^{+0.011}_{-0.006}$ &  4.515$^{+0.024}_{-0.009}$ &   -0.01$^{+0.15}_{-0.05}$ \\
HD 109573     & 9641$\pm$61      &  2.14$\pm$0.09           &  1.694$\pm$0.011         &  22.3$\pm$0.5            &  4.311$\pm$0.022         &    0.09$\pm$0.23 \\
HD 121617     & 9029$\pm$34      &  1.901$\pm$0.009         &  1.526$\pm$0.011         &  13.94$\pm$0.13          &  4.350$\pm$0.005         &    0.024$\pm$0.020 \\
HD 131488     & 8713$\pm$30      &  1.804$\pm$0.016         &  1.493$\pm$0.015         &  11.56$\pm$0.17          &  4.346$\pm$0.008         &   -0.00$\pm$0.03 \\
HD 131835     & 8284$\pm$28      &  1.698$\pm$0.013         &  1.464$\pm$0.010         &  9.09$\pm$0.10           &  4.337$\pm$0.007         &   -0.02$\pm$0.03 \\
HD 145560     & 6467$\pm$19      &  1.35$\pm$0.08           &  1.422$\pm$0.016         &  3.18$\pm$0.04           &  4.26$\pm$0.03           &    0.09$\pm$0.15 \\
HD 161868     & 8658$\pm$56      &  2.11$\pm$0.07           &  2.075$\pm$0.027         &  21.8$\pm$0.4            &  4.129$\pm$0.021         &    0.29$\pm$0.17 \\
HD 170773     & 6667$\pm$28      &  1.40$\pm$0.05           &  1.396$\pm$0.021         &  3.47$\pm$0.05           &  4.293$\pm$0.028         &    0.17$\pm$0.25 \\
HD 197481        & 3615$\pm$14      &  0.614$\pm$0.026         &  0.805$\pm$0.005         &  0.0998$\pm$0.0005       &  4.414$\pm$0.023         &   -0.15$\pm$0.08 \\
HD 206893     & 6571$\pm$27      &  1.329$\pm$0.011         &  1.281$\pm$0.011         &  2.757$\pm$0.007         &  4.346$\pm$0.004         &    0.14$\pm$0.05 \\
TYC9340-437-1 & 4091$\pm$12      &  0.753$\pm$0.009         &  0.855$\pm$0.006         &  0.1844$\pm$0.001        &  4.451$\pm$0.010         &    0.25$\pm$0.06 \\
HD 218396       & 7366$^{+31}_{-33}$ &  1.495$^{+0.011}_{-0.034}$ &  1.414$^{+0.010}_{-0.012}$ &  5.299$^{+0.020}_{-0.040}$ &  4.312$^{+0.010}_{-0.018}$ &   -0.05$^{+0.05}_{-0.16}$ \\
    \hline
    \end{tabular}

    \tablefoot{ Effective temperature $T_{\rm eff}$, mass $M_{\star}$, radius $R_{\star}$, luminosity $L_{\star}$, surface gravity $\log(g)$, and iron-to-hydrogen abundance [Fe/H], all output parameters from the Bayesian inference code of \citet{delburgo2016,delburgo2018}.}
    
    \label{tab:stellarparam}
\end{table*}

\section{ALMA measured fluxes}
\label{app:fluxes}

\begin{table*}
    \caption{Phase centres and integrated fluxes of the 24 systems analysed in ARKS derived from the primary beam corrected images. }
    \centering
    \begin{tabular}{l c c c c c c c c}
    \hline
    \hline
      Name  & Date    & RA & Dec & $\Delta$RA$_{\rm G}$ & $\Delta$Dec$_{\rm G}$ & Wavelength & Flux  & Aperture size   \\
            & [d/m/y] & [h:m:s] &   [d:m:s]  & [mas] & [mas] &  [mm]      & [mJy] & [au (\arcsec)] \\
     \hline
HD 9672 & 17/06/2024  & 01:34:37.9394 & -15:40:34.967 & 1  & 0  & 0.89 & $13.9\pm 1.4$ & 411 (7.2) \\
HD 10647 & 08/05/2024  & 01:42:29.7722 & -53:44:29.508 & 1  & 0  & 0.89 & $14.8\pm 1.5$ & 197 (11.3) \\
HD 14055 & 26/07/2024  & 02:17:18.9573 & +33:50:48.612 & 17  & 10  & 0.89 & $8.1\pm 0.9$ & 333 (9.3) \\
HD 15115 & 09/05/2023  & 02:26:16.3839 & +06:17:32.012 & -1  & 1  & 0.89 & $4.8\pm 0.5$ & 144 (2.9) \\
HD 15257 & 19/10/2022  & 02:28:09.9531 & +29:40:07.632 & -3  & -3  & 0.89 & $3.8\pm 0.5$ & 352 (7.2) \\
HD 32297 & 23/09/2023  & 05:02:27.4455 & +07:27:39.119 & -1  & 2  & 0.89 & $10.8\pm 1.1$ & 308 (2.4) \\
HD 39060 & 08/08/2015  & 05:47:17.0931 & -51:03:58.137 & 67  & 46  & 1.33 & $18.6\pm 1.9$ & 194 (9.9) \\
HD 61005 & 11/05/2023  & 07:35:47.3591 & -32:12:12.298 & 0  & 1  & 0.89 & $14.8\pm 1.5$ & 135 (3.7) \\
HD 76582 & 16/10/2022  & 08:57:35.2957 & +15:34:53.094 & 0  & 0  & 0.89 & $8.1\pm 0.9$ & 489 (10.0) \\
HD 84870 & 24/12/2022  & 09:49:02.7913 & +34:05:06.134 & 0  & 0  & 0.89 & $2.6\pm 0.3$ & 413 (4.7) \\
HD 92945 & 17/12/2016  & 10:43:27.9966 & -29:03:52.304 & -8  & 5  & 0.86 & $11.1\pm 1.1$ & 156 (7.3) \\
HD 95086 & 09/01/2023  & 10:57:02.8496 & -68:40:02.161 & 0  & 0  & 0.89 & $9.3\pm 1.0$ & 467 (5.4) \\
HD 107146 & 03/05/2021  & 12:19:06.2413 & +16:32:50.707 & -1  & -1  & 1.00 & $22.1\pm 2.2$ & 182 (6.6) \\
HD 109573 & 04/06/2023  & 12:36:00.9175 & -39:52:10.778 & 0  & -1  & 0.89 & $13.8\pm 1.4$ & 116 (1.6) \\
HD 121617 & 11/05/2023  & 13:57:41.0610 & -47:00:34.792 & 0  & 0  & 0.89 & $4.2\pm 0.4$ & 157 (1.3) \\
HD 131488 & 27/06/2023  & 14:55:07.9902 & -41:07:13.925 & 0  & 0  & 0.89 & $7.4\pm 0.8$ & 364 (2.4) \\
HD 131835 & 30/04/2023  & 14:56:54.4230 & -35:41:44.244 & 0  & 0  & 0.89 & $6.6\pm 0.7$ & 298 (2.3) \\
HD 145560 & 19/05/2023  & 16:13:34.2891 & -45:49:04.446 & 0  & 0  & 0.89 & $4.6\pm 0.5$ & 270 (2.2) \\
HD 161868 & 01/10/2022  & 17:47:53.5203 & +02:42:24.505 & 27  & 4  & 0.89 & $7.5\pm 0.8$ & 283 (9.5) \\
HD 170773 & 23/01/2023  & 18:33:01.0908 & -39:53:33.111 & 0  & 0  & 0.89 & $14.2\pm 1.5$ & 252 (6.8) \\
HD 197481 & 18/08/2014  & 20:45:09.8511 & -31:20:32.517 & 5  & -19  & 1.35 & $4.7\pm 0.5$ & 52 (5.3) \\
HD 206893 & 30/08/2018  & 21:45:22.0239 & -12:47:00.064 & 15  & -6  & 1.35 & $1.0\pm 0.1$ & 200 (4.9) \\
TYC 9340-437-1 & 02/10/2022  & 22:42:49.3897 & -71:42:22.415 & 0  & 0  & 0.89 & $9.5\pm 1.0$ & 227 (6.2) \\
HD 218396 & 29/05/2018  & 23:07:28.8587 & +21:08:02.399 & 15  & -2  & 0.88 & $7.1\pm 0.8$ & 412 (10.1) \\
    \hline
    \end{tabular}

    \tablefoot{Columns 2--6 display the date and the phase centre coordinates of the reference execution block in the ICRS frame used to align the observations during the correction step, and the offset relative to the predicted stellar position according to Gaia DR3 \citep[the RA offset has been multiplied by $\cos(\mathrm{Dec})$,][]{gaiadr3}. Columns 7--9 show the wavelength, integrated flux, and aperture size. For systems with submillimetre galaxies, we used the subtracted images. The flux errors include a nominal 10\% absolute flux calibration error added in quadrature to the statistical error that arises from the image rms. The aperture size corresponds to the semi-major axis of an elliptical mask used to measure the flux.}
    
    \label{tab:fluxes}
\end{table*}

We derive integrated fluxes for each system as follows. We use the corresponding clean image produced with a robust parameter of 2.0, primary beam corrected, and with SMGs subtracted. We then create a series of concentric elliptical masks with increasing semi-major axes. These masks match the ellipticity of the discs given their inclination (as constrained by the MCMC fits) and also take into account the beam size and a scale height as large as 10\% of the disc radius. We integrate the intensity within these masks to derive an integrated flux as a function of aperture size and the flux uncertainty taking into account the beam size and image rms, which increase towards the edge of the images due to the primary beam response. This curve is then used to determine the final aperture size, set as the radius at which the minimum flux (defined as flux minus 3 times the uncertainty) stops increasing. Finally, we add in quadrature a 10\% flux calibration error to the flux uncertainty. The wavelength, measured fluxes and aperture sizes are summarised in Table~\ref{tab:fluxes}.

We find that all the measured fluxes of the 24 systems in ARKS are within 3$\sigma$ of those expected from or measured by REASONS (taking into account REASONS and ARKS uncertainties). The only fluxes that have more than a 30\% difference are those of HD~15257 and HD~84870, which ARKS measures to be ${\sim}$50\% fainter but still within 2$\sigma$. The measured fluxes are also within $15\%$ of the flux of our best-fit simple models for all systems, except for HD~131488. For this system, the model flux is 5.5~mJy instead of the 7.4~mJy measured from the image. The lower model flux could be due to it missing the emission interior and exterior of the main ring, as it was modelled as a Gaussian \citep{rad_arks}. Similarly, the smaller differences for the other systems can be explained by the model assumptions of the surface density profile being a Gaussian.

\section{Background sources}
\label{app:smg}

\begin{figure*}
    \centering
    \includegraphics[trim=0.0cm 0.0cm 0.0cm 0.0cm,
    clip=true, width=1.0\textwidth]{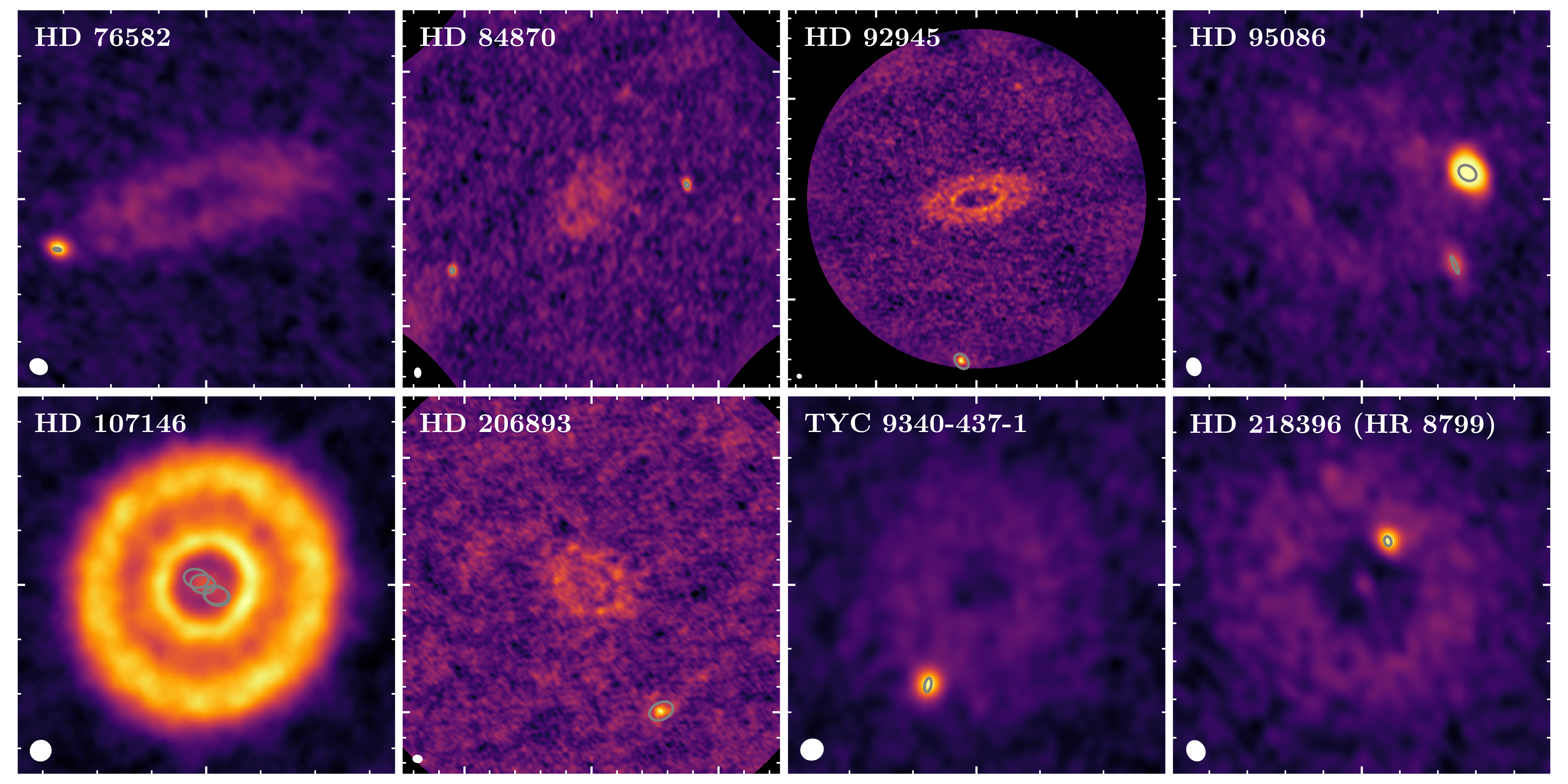}
    \caption{ARKS continuum clean images of the eight systems in the sample with background sources before their subtraction.
    The data were imaged with a robust value of 2 and the white ellipse in the bottom left corner represents the corresponding beam. The grey ellipses represent the best-fit location and size (FWHM) of the SMGs at the different epochs of observations. For HD~95086 the image has a non-linear scale to increase the dynamic range and to visualise both SMGs. The white minor and major ticks in all panels are spaced by 2\arcsec\ and 10\arcsec, respectively.
    }
    \label{fig:smg}
\end{figure*}

\begin{table*}
    \caption{Submillimetre galaxies best-fit parameters.}
    \centering
    \begin{tabular}{l c c c c c c c c }
    \hline
    \hline 
    Name  & ID & RA & Dec & $\lambda$ & SMG Flux & $\sigma_{\rm maj}$ & $\sigma_{\rm min}$ & PA  \\
          &    &  [h:m:s]  &  [d:m:s]  & [mm] & [mJy] & [\arcsec]         & [\arcsec]  & [\degr] \\
    \hline
HD 76582 & 1 & 08:57:35.73 & +15:34:50.97 & 0.89 & $0.74 \pm 0.04$ & $0.131 \pm 0.018$ & $0.052 \pm 0.008$ & $80 \pm 16$ \\
HD 84870 & 1 & 09:49:02.19 & +34:05:07.24 & 0.89 & $0.34 \pm 0.03$ & $0.189 \pm 0.042$ & $0.108 \pm 0.017$ & $19 \pm 18$ \\
HD 84870 & 2 & 09:49:03.67 & +34:05:00.51 & 0.89 & $0.75 \pm 0.08$ & $0.187 \pm 0.045$ & $0.113 \pm 0.023$ & $4 \pm 48$ \\
HD 92945 & 1 & 10:43:28.11 & -29:04:08.49 & 0.86 & $6.60 \pm 0.60$ & $0.708 \pm 0.140$ & $0.499 \pm 0.171$ & $41 \pm 34$ \\
HD 95086 & 1 & 10:57:02.34 & -68:40:01.48 & 0.89 & $3.84 \pm 0.06$ & $0.210 \pm 0.004$ & $0.157 \pm 0.003$ & $57 \pm 3$ \\
HD 95086 & 2 & 10:57:02.40 & -68:40:03.89 & 0.89 & $0.43 \pm 0.04$ & $0.209 \pm 0.034$ & $0.043 \pm 0.027$ & $23 \pm 6$ \\
HD 107146 & 1 & 12:19:06.27 & +16:32:50.95 & 0.87 & $0.36 \pm 0.04$ & $0.388 \pm 0.047$ & $0.285 \pm 0.032$ & $78 \pm 15$ \\
HD 206893 & 1 & 21:45:21.65 & -12:47:09.97 & 1.35 & $0.34 \pm 0.04$ & $0.832 \pm 0.112$ & $0.567 \pm 0.093$ & $115 \pm 13$ \\
TYC 9340-437-1 & 1 & 22:42:49.71 & -71:42:25.56 & 0.89 & $0.87 \pm 0.04$ & $0.191 \pm 0.022$ & $0.085 \pm 0.023$ & $-12 \pm 9$ \\
HD 218396 & 1 & 23:07:28.76 & +21:08:04.70 & 0.88 & $0.30 \pm 0.02$ & $0.216 \pm 0.036$ & $0.148 \pm 0.035$ & $17 \pm 27$ \\
    \hline 
    \end{tabular}
    \tablefoot{Best-fit parameters were derived from our MCMC using the median and the 16th and 84th percentiles to estimate the uncertainties. Columns 7--9 show the standard deviation of the major axis, the standard deviation of the minor axis, and the position angle of the deconvolved elliptical Gaussian.}
    
    \label{tab:smg}
\end{table*}

The compact, bright sources detected in the field of view of eight of our targets were identified as background submillimetre galaxies. We are confident that these sources are not related to the discs for the following reasons. First, those in HD~84870, HD~92945, and HD~206893 images have projected distances that would put these objects at hundreds of au. Second, the spectral indices derived by previous studies in the three systems with multi-band data are larger than 3 and thus consistent with SMGs \citep[this is the case of HD~95086, HD~107146, and HD~218396,][]{Booth2019, Marino2021, Faramaz2021}. Third, those with archival and ARKS data spanning years show negligible proper motion \citep[this is the case for HD~76582, HD~95086, HD~107146, and HD~218396,][]{Matra2025, Zapata2018, Imaz-Blanco2023, Faramaz2021}. Finally, the source detected near TYC~9340-437-1 has recently been detected with JWST/NIRCam at 4$\mu$m at a location that indicates it is not co-moving with the system (Zhang et al. in prep).

Table~\ref{tab:smg} presents the best-fit parameters of the SMGs identified in 8 of our targets. Figure~\ref{fig:smg} shows the clean continuum images without the SMGs subtracted and using Briggs weighting with a robust value of 2.0. The location and best-fit morphology of the SMGs are highlighted with grey ellipses; HD107146's observations span multiple years, and thus, the SMG's location changed relative to the star. We note that for two systems, the subtractions were not optimal and left significant residuals: HD~95086 and HD~107146.

For HD~107146, the fit converged to a total SMG flux of 0.36 mJy at 0.87~mm, which is significantly lower than the 0.79~mJy estimated at a similar wavelength in a previous study \citep{Marino2021}. The lower retrieved flux is likely due to the choice of disc model. \cite{Marino2021} fit a more flexible disc model to the data that included an inner edge that was parametrised such that it could be smooth or sharp independently of the outer edge and gap. Here, instead, we used a simpler double Gaussian to fit the disc, which effectively imposes a smooth disc inner edge. Therefore, it is likely that our disc model overpredicted the disc surface brightness just interior to the disc where the SMG is found in some epochs. This could explain why the fit converged to a lower SMG flux. This lower SMG flux is likely the reason why the radial profile presented in \cite{rad_arks} shows emission at ${\sim}20$~au that is not present in previous analyses \citep{Marino2021, Imaz-Blanco2023}.

For HD~95086, we fit two components to the two brightest features in the data that are found ${\sim}3\arcsec$ to the west. Our model converged to SMG fluxes of 3.8 and 0.4 mJy, which are consistent with the fluxes derived by \cite{Zapata2018} using lower-resolution observations. However, the SMG subtracted image in Fig.~\ref{fig:gallery} still displays excess emission towards the west and near the brightest SMG, which is likely additional background emission. Recent JWST/MIRI observations support the idea of more extended and complex background emission \citep{Malin2024}. While the 23~$\mu$m image shows two background sources that match the SMG positions \citep[labelled "g" and "g2" in][]{Malin2024}, at 11~$\mu$m g is shifted towards the NE and it appears more extended than in the ALMA images. We thus conclude that there may be a third extended source in the ALMA images (labelled g at 11~$\mu$m) or additional extended emission associated with the brightest SMG.

\section{Stellar variability}
\label{sec:variability}

\begin{figure}
    \centering
    \includegraphics[trim=0.0cm 0.0cm 0.0cm 0.0cm,
    clip=true, width=1.0\columnwidth]{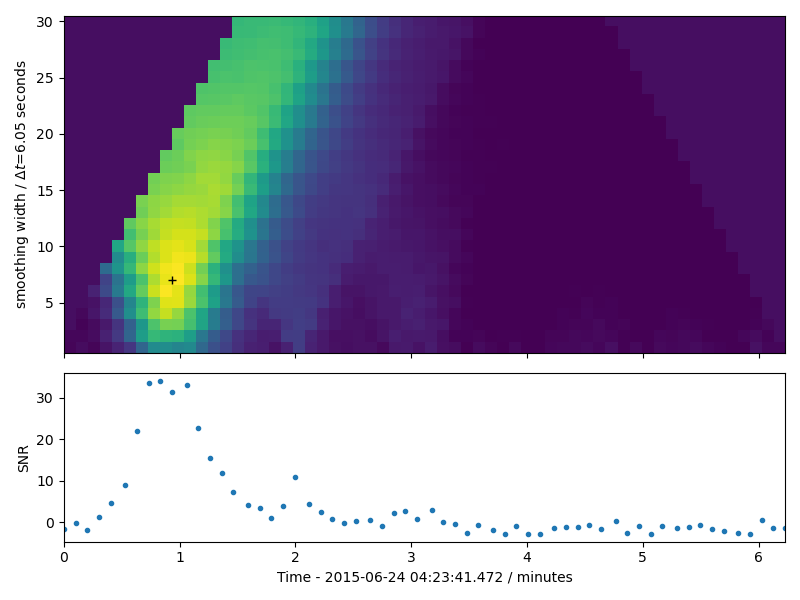}
    \caption{Example of the stellar variability search output for scan 27 from the 2015 AU~Mic data. The lower panel shows the S/N vs. time for a single scan (27), with one large flare visible. The colour scale in the upper panel shows the mean-filtered data from the lower panel using a window of the width given on the y-axis to increase the S/N for flares with a duration greater than 6\,seconds. The flare is mostly strongly detected for a window of $7 \times 6$\,s, as indicated by the plus sign (+)  in the upper panel.}
    \label{fig:aumic_var}
\end{figure}

We performed a search for stellar variability on all data, using spectrally averaged MS files with no time averaging.\footnote{The code is publicly available at \href{https://github.com/drgmk/alma_var}{https://github.com/drgmk/alma\_var}.} In most cases the star is not detected by ALMA, so this search is primarily for large positive variations in flux that make the star temporarily visible, i.e. flares. Each MS was further averaged down to one channel per spectral window to reduce the data volume, and a model based on a clean image from the full MS was subtracted to produce the MS to be searched, which in the absence of variability is simply Gaussian noise (which is verified with a Shaprio-Wilk test). Each model-subtracted MS was then split into individual files for every scan (duration $\sim$5\,minutes), for which there is typically one measurement every 6\,seconds (i.e.\ 50 time steps per scan). These scans were searched in visibility space for variability using a matched filter for a point source \citep[e.g.][]{Loomis2018}, assuming equal flux in each spectral window, at the locations of any Gaia DR3 sources within the primary beam (which includes our target star in each case). The matched filter for each $u,v$ point is therefore a complex number with unit length that is rotated according to the search location in sky coordinates. The filter spans only a single time point (i.e. we are not trying to match the time-dependent decay of a flare), so the output is an S/N for each time step. These S/N's were then time averaged with a mean filter for windows up to half the scan length, to improve the S/N for variability on timescales longer than an individual time step. S/N values greater than 4 for any window length were flagged for visual inspection. An example of the output for a single scan is shown in Fig.\,\ref{fig:aumic_var}. This method recovered the flare seen for HD~197481 \citep{Daley2019}, and this scan was removed from the data. We did not detect significant variability for any other targets. It is possible that this method could miss small flares if the average stellar flux were biased upwards by flares, as the S/N's would be biased downwards (e.g. the post-flare S/N values in Fig.\,\ref{fig:aumic_var} are below zero). We do not see any evidence that marginally significant flares were missed in this way, but an improved method could iteratively exclude scans with S/N's significantly different from zero in creating the model that is subtracted from the original MS. The upper limits set on the flux of any flares in each case varies, depending on the point source sensitivity of the observations and the stellar flux. For example, for HD~145560 the stellar flux is expected to be about 4\,mJy, but the rms is 9\,mJy, so in that case any flares would need to be 7 times the stellar flux to have been detected. Flares stronger than this could also have been missed if their timescale were significantly shorter than 6\,seconds.

\section{Multiplicity}
\label{app:companions}

To understand the architecture of the 24 systems studied in ARKS, we performed a literature search for any information on stellar or substellar companions. These can usually be detected through radial velocities with spectroscopy \citep[e.g.][]{Latham2002, Fernandez2017},\footnote{Spectroscopic binaries are abbreviated as SB1s when only one component is detected, SB2 when two are detected, and so on.} through astrometry including astrometric accelerations or  proper motion anomalies \citep[PMa,][]{Kervella2019, Brandt2021pma}, astrometric excess noise \citep[e.g. as measured by Gaia's RUWE parameter,][]{Belokurov2020}, or resolving a companion using adaptive optics, speckle imaging or interferometry \citep{JimenezEsteban2019, Dallant2023}.

Our literature search focused on studies with electronic catalogues available, which we queried using \texttt{astroquery} within \textsc{astropy} \citep{Ginsburg2019, astropy:2022} and complemented with a simple visual inspection of the fields in Gaia Data Release 3 \citep{gaiadr3}, in the images of the Two Micron All Sky Survey (2MASS) in the J, H, and K bandpasses \citep{Skrutskie2006}, in the Sloan Digital Sky Survey (SDSS) images in the r and g bandpass \citep{Blanton2017}, and in the Digitized Sky Survey (DSS) image \citep{DSS2020}. Table \ref{tab:Catalogs1a} presents the queried catalogues containing ARKS targets.

\begin{table}
\centering
\caption{Queried catalogues containing ARKS targets.}
\centering
\begin{tabular}{ll}
\hline
\hline
Reference & Vizier Catalog ID \\ \hline
$^{R01}$\cite{Gratton2024} & Not in VizieR (as of 2024)\\
$^{R02}$\cite{Kervella2022}& J/A+A/657/A7 \\
$^{R04}$\cite{Gonzalez-Payo2024}& J/A+A/689/A302 \\
$^{R05}$\cite{gaiadr3}& I/355/gaiadr3 \\
$^{R07}$\cite{Elliott2016} & J/A+A/590/A13 \\
$^{R08}$\cite{Biller2013}& J/ApJ/777/160 \\
$^{R09}$\cite{Baron2018}& J/AJ/156/137 \\
$^{R12}$\cite{Elliott2014} & J/A+A/568/A26 \\
$^{R13}$\cite{Zuniga-Fernandez2021} & J/A+A/645/A30 \\
$^{R17}$\cite{Brandt2021pma}& J/ApJS/254/42 \\
$^{R27}$\cite{Elliott2015} & J/A+A/580/A88 \\
$^{R37}$\cite{Tokovinin2018}& J/ApJS/235/6 \\
$^{R38}$\cite{Mason2001}& B/wds\\
\cite{Cifuentes2025} & Not in VizieR (as of 2024)\\
\hline
$^{R03}$\cite{Butler2006}&J/ApJ/646/505\\
$^{R06}$\cite{Vigan2012}&J/A+A/544/A9\\
$^{R10}$\cite{Rosa2014}&J/MNRAS/437/1216\\
$^{R11}$\cite{Lagrange2020}&Not in VizieR (as of 2024)\\
$^{R14}$\cite{Chauvin2015}&Not in VizieR (as of 2024)\\
$^{R15}$\cite{Rodriguez2012}&J/ApJ/745/147\\
$^{R16}$\cite{Rameau2013}&Not in VizieR (as of 2024)\\
$^{R18}$\cite{Jura1993}&Not in VizieR (as of 2024)\\
$^{R19}$\cite{Kastner2008}&Not in VizieR (as of 2024)\\
$^{R20}$\cite{Zakhozhay2022}&J/A+A/667/A63\\
$^{R21}$\cite{Janson2013}&Not in VizieR (as of 2024)\\
$^{R22}$\cite{Janson2013_2}&Not in VizieR (as of 2024)\\
$^{R23}$\cite{Plavchan2020}&Not in VizieR (as of 2024)\\
$^{R24}$\cite{Martioli2021}&Not in VizieR (as of 2024)\\
$^{R25}$\cite{Grandjean2019}&Not in VizieR (as of 2024)\\
$^{R26}$\cite{Zurlo2022}&Not in VizieR (as of 2024)\\
$^{R29}$\cite{Bertini2023}& J/A+A/671/L2\\
$^{R30}$\cite{Lagrange2009}&Not in VizieR (as of 2024)\\
$^{R31}$\cite{Lagrange2019}&Not in VizieR (as of 2024)\\
$^{R33}$\cite{Nielsen2019}&J/AJ/158/13 \\
$^{R34}$\cite{Ducourant2014}&Not in VizieR (as of 2024)\\
$^{R36}$\cite{Shaya2011}&J/ApJS/192/2 \\
$^{R39}$\cite{Hinkley2023}&Not in VizieR (as of 2024)\\
$^{R40}$\cite{Milli2017}&Not in VizieR (as of 2024)\\
$^{R41}$\cite{Marois2008}&Not in VizieR (as of 2024)\\
$^{R42}$\cite{Marois2010}&Not in VizieR (as of 2024)\\
\hline
$^{R28}$\cite{Marmier2013}& \\
$^{R32}$\cite{Chauvin2018}& \\
$^{R35}$\cite{Mallorquin2024}& \\
\hline
\end{tabular}
\tablefoot{The first part of the table lists the main catalogues queried and the second part those queried at a second instance since they were referenced by the main catalogues. The third part lists the references that were used in the multiplicity study, due to their occurrence in Table \ref{tab:sample}.}
\label{tab:Catalogs1a}
\end{table}

In addition to the catalogues listed in Table~\ref{tab:Catalogs1a}, we searched the following catalogues but did not recover any entry in common with the ARKS sample:
\begin{itemize}
    \item Gaia DR3: Gaia Data Release 3: Non-single stars (2022). Vizier: "I/357" \cite{GDR3_NonSingleStars2022};
\item Gaia EDR3 nearby accelerating stars: A Catalogue of Nearby Accelerating Star Candidates in Gaia DR3 (2023). Vizier: "J/AJ/165/193" \cite{Whiting2023};
\item The Gaia ultracool dwarf sample - V: the ultracool dwarf companion catalogue. No VizieR entry at the time of writing. Tables extracted from the body of the paper \cite{Baig2024};
\item Combining HIPPARCOS and Gaia data for the study of binaries: The BINARYS tool. No VizieR entry at the time of writing. Tables extracted from the body of the paper \cite{Leclerc2023};
\item Spectroscopic Follow-up of Gaia Exoplanet Candidates: Impostor Binary Stars Invade the Gaia DR3 Astrometric Exoplanet Candidates. No Vizier entry at the time of writing. Tables in the paper \cite{Marcussen2023};
\item Binarity and beyond in A stars - I. Survey description and first results of VLTI/GRAVITY observations of VAST targets with high Gaia-Hipparcos accelerations. No VizieR entry at the time of writing. Tables extracted from the body of the paper \cite{Waisberg2023};
\item Multiples among B-stars in Sco-Cen. Vizier: "J/A+A/678/A93", \cite{Gratton2023};
\item Combining Gaia and GRAVITY: Characterising five new directly detected substellar companions. No VizieR entry at the time of writing. Tables extracted from the body of the paper \cite{Winterhalder2024}.
\end{itemize}

\subsection{Multiplicity results}

A total of eight systems were identified to have confirmed companions, exhibiting a wide range of mass ratios and separations (see Table~\ref{tab:companions} for details). Four of these eight systems (HD~10647, HD~76582, HD~109573 and HD~197481) have stellar companions at wide separations from ${\sim}500-4\times10^{4}$~au. One of these eight systems has a brown dwarf companion (HD~206893) interior to the belt detected via RV, PMa and direct imaging. Six of the eight systems (HD~10647, HD~39060, HD~95086, HD~197481, HD~206893 and HD~218396) contain companions in the planetary mass regime interior to their belts, at separations that are no more than 50\% the inner edge of the belts.

\begin{table*}
\centering
\caption{Multiplicity status per system and range of separations. 
}
\begin{tabular}{l c c c }
\hline
\hline
 & \multicolumn{3}{c}{Status per range of separation \footnotemark[1]} \\
Name & Close & Intermediate & Wide \\ 
& $\lesssim$ 1 au & $\sim$ few au - tens au & $\gtrsim$ 100 au\\ 
\hline 
HD 9672                  & ? & N & ?\\ 
                         & RV $^{R01}$ & AO+RUWE $^{R01}$ & \\ 
HD 10647                 & Y? & Y & Y \\
                         & PMa $^{R02}$ & PMa+CPM$^{R02}$, RV$^{R03}$ & PMa+CPM $^{R02}$, CPM$^{R04}$ \\
HD 14055                 & Y? & N & N \\
                         & RUWE $^{R05}$ & AO $^{R06}$ & CPM $^{R07}$\\
HD 15115                 & ? & N & N\\
                         & RV $^{R01}$ & AO$^{R06}$, RUWE$^{R01}$, CPM$^{R08}$ & CPM $^{R09}$\\
HD 15257                 & N? & N & N \\
                         & PMa+RUWE $^{R02}$ & AO$^{RTW3}$ & CPM $^{R10}$ \\ 
HD 32297                 & N? & N & ?\\
                         & PMa+RUWE $^{R02}$ & AO$^{RTW1, RTW2}$ & ?\\
HD 39060                 & N & Y & N \\
                         & RV$^{R11}$ & AO+RV$^{R11}$ & CPM$^{R09}$ \\
HD 61005                 & N & N & N \\
                         & RV $^{R12}$$^{R13}$ & AO+CPM $^{R14}$ & CPM $^{R07}$$^{R09}$\\
HD 76582 & N? & N & Y \\
                         & PMa $^{R02}$ & AO$^{RTW3}$ & CPM $^{R02}$$^{R36}$\\ 
HD 84870                 & N? & N & N\\
                         & PMa+RUWE $^{R02}$ & AO$^{R15}$ & AO $^{R15}$\\
HD 92945 & Y? & N & ? \\
                         & PMa $^{R02}$ & AO$^{RTW1, RTW2}$ & \\ 
HD 95086                 & ? & Y & N  \\
                         & ? & AO $^{R16}$ & AO $^{R16}$ \\
HD 107146                & Y? & N & ?\\
                         & PMa $^{R02}$$^{R17}$ & AO$^{RTW1, RTW2}$ & \\ 
HD 109573                & A: ?, B: N & N & Y \\
                         & B: RV$^{R13}$ & AO$^{RTW1}$ & CPM$^{R18}$$^{R07}$$^{R19}$ \\
HD 121617                & ? & ? & ?\\
\\
HD 131488                & ? & N & ?\\
                & & AO$^{RTW1, RTW2}$ &\\
HD 131835                & ? & N & ?\\
                & & AO$^{RTW2}$ & \\
HD 145560                & ? & N & N\\
                         & RV$^{R20}$ & AO$^{R21}$ & AO$^{R21}$\\
HD 161868                & Y? & N & N \\
                         & RUWE+PMa $^{R02}$ & AO$^{RTW3}$ & AO $^{R22}$ \\ 
HD 170773                & ? & N & ?\\
                & & AO$^{RTW1, RTW2}$ & \\
HD 197481                & Y & N & Y  \\
                         & T$^{R23}$$^{R24}$ & AO$^{RTW1, RTW2}$ & AO$^{R12}$ \\
HD 206893                & N & Y & N \\
                         & RV$^{R25}$ & AO$^{R25}$ & AO$^{R25}$ \\ 
HD 218396                & Y? & Y & Y? N \\
                         & RUWE$^{R02}$ & AO$^{R26}$ & Y?: CPM$^{R02}$, CPM$^{R07}$$^{R09}$ \\ 
TYC 9340-437-1           & N & N & N \\
                         & RV$^{R13}$ & AO$^{RTW1, RTW2}$ & AO$^{R27}$, CPM$^{R07}$\\\hline

\end{tabular}

\tablefoot{References$^{RX}$ for the quoted techniques as in Table~\ref{tab:Catalogs1a}. Flags:
        ? = no data available and/or no conclusion on the (non)presence of a companion. 
        Y = existing surveys report detected (gravitationally bound) stellar and/or substellar companions.
        Y? = RUWE and/or PMa and/or $\chi^2$ and/or searches for CPM candidates indicate possible companions.
        N = existing surveys probing this separation range report non-detections $\rightarrow$ with (high) probability there are no companions.
         N? = values exist of PMa and/or RUWE that do not show signs of multiplicity
        Techniques: RV = Radial Velocity; AO = Adaptive Optics; PMa = Proper Motion Anomaly (mostly between HIPPARCOS and Gaia); RUWE = Renormalized Unit Weight Error (from Gaia); CPM = Common Proper Motion; T = Transit.
        Flags with the reference `$^{RTW}$' correspond to contrast limits obtained homogeneously with the data available in the SPHERE data centre, stellar properties from Table \ref{tab:sample}, and the predictions in mass from the \cite{Baraffe2015} model using broad-band H filter (and if not available $H\_D2$ and $H\_D3$ filter) observations: we separated the outcomes into  three regimes:
        RTW1: Objects for which we can discard stellar companions down to 3au;
        RTW2: Objects for which we can discard companions with masses $>$ 13 Mjup as close as 5au;
        RTW3: Objects for which we can discard stellar companions down to $\sim$10au.

}

\label{tab:ConfRejSep}
\end{table*}

\begin{table*}
\caption{Architecture of multiple system and properties of the companion.}

\begin{adjustbox}{max width=0.98\textwidth}
\begin{threeparttable}
\begin{tabular}{l l c l l l l l}
\hline
\hline
&  \multicolumn{4}{c}{Properties of confirmed multiple systems \footnotemark[1]}\\
Name & $M_{P}$ & Config. & Sep & $M_{\rm c}$ & $\Delta M_{\rm c}$ & $M_{\rm c}$ & ($R-\Delta R/2$)--($R+\Delta R/2$) \footnotemark[2] \\
& [$M_{\odot}$] & &[au]& [$M_{\rm Jup}$] & [$M_{\rm Jup}$] & [$M_{\odot}$] & [au]\\ 
\hline 
HD 10647                 & 1.12 & Ab$^{R03}$ & 2.0$^{R28}$ & 0.93$^{R03}$ & 0.18 & & 71--136\\
& & B $^{R02}$ & $3.7\times10^{4}$ $^{R02,R29}$ & & & [0.2,0.4]$^{R29}$ &\\
HD 39060                 & 1.72 & b$^{R30}$ & 10.3$^{R17}$ & 10--11$^{R11}$ & & & 56--153\\
& & c$^{R31}$ & 2.7$^{R17}$ & 7.8$^{R11}$ & 0.4 & &\\
HD 76582\footnotemark[3] & 1.61 & B $^{R02}$ & $4.8\times 10^{4}$ $^{R02}$ & ? & & & 116--292\\
HD 95086                 & 1.54 & b$^{R16}$ & 52$^{R32}$ & 2.6$^{R33}$ & 0.4 & & 110--284\\
HD 109573                & 2.14 & B$^{R18}$ & 490-560$^{R18,R02}$ & & & 0.37 $^{R34}$ & 72--79\\
& & C$^{R18}$ & 1.2--1.4$\times10^{4}$ $^{R07,R19,R02}$ & & & 0.1 $^{R34}$ &\\
HD 197481\footnotemark[4]                & 0.61 & Ab$^{R23}$ & 0.07$^{R35}$  & 0.028$^{R35}$ & 0.009 & & 27--40\\
& & Ac$^{R23}$ & 0.12$^{R35}$ & 0.04$^{R35}$ & 0.02 & &\\
& & AB = AT Mic$^{R36}$ & 4.0--4.5$\times 10^4$ $^{R37,R38,R36}$ & & & 0.37 $^{R37}$, 0.36$^{R36}$\\
& & BC$^{R37}$ & 20--24$^{R02}$ & & & 0.20 $^{R37}$\\
HD 206893                & 1.33 & Ab$^{R25}$ & 3.5$^{R39}$ & 12.7$^{R39}$ & 1.2 & & 47--163\\
& & B$^{R40}$ & 9.7$^{R39}$ & 28.0$^{R39}$ & 2.2 & & \\
HD 218396                & 1.50 & b $^{R41}$ & 72$^{R26}$ & 6$^{R26}$ & & & 149--331\\
& & c $^{R41}$ & 41$^{R26}$ & 8$^{R26}$\\
& & d $^{R41}$ & 27$^{R26}$ & 9$^{R26}$\\
& & e $^{R42}$ & 16$^{R26}$ & 8 $^{R26}$\\
\hline
\end{tabular}
\tablefoot{
References$^{RX}$ for the quoted values are provided as in Table~\ref{tab:Catalogs1a}. Column 8 shows the disc extent.

    $^{1}M_P$: Primary Mass: same as in Table \ref{tab:stellarparam}: estimated as described in Appendix \ref{sec:star}; Config.: Architecture of the system: lower case letters indicate planetary companions, capital letters indicate stellar or brown dwarf companions; Sep: semi-major axis for those companions with orbital constraints (Sep$<100$~au) or separation for those on wider orbits (homogenised to angular units with parallaxes from Gaia DR3); $M_{\rm c}$: Companion Mass: for substellar companions the units are given in Jupiter masses, while for stellar companions the mass is given in solar masses. If provided in the referenced literature, the uncertainties are given in $\Delta M_{\rm c}$
    $^{2}$Approximated disc extent. Values of $R$ and $\Delta R$ are taken from Table \ref{tab:mcmc_results}
    $^{3}$ HD 76582 and HD 76543 reported by \cite{Kervella2022} and \cite{Shaya2011} as a common proper motion pair. In addition to the finding from \cite{gaiadr3}, the two stars are at the same distance and have radial velocities that are consistent within the errors. No mass is reported for the companion; instead, a placeholder (‘?’) is used. This is because \cite{Kervella2022} does not provide a mass estimate, and the masses for both components reported in other studies, such as \cite{Shaya2011} and \cite{Rosa2014}, are inconsistent with the primary star adopted here and with each other.
    $^{4}$ Hierarchical stellar triple system: The primary star HD 197481 (AU Mic) hosts  two confirmed planets;  and there are hints for two additional planets (\citealt{Donati2023}). It has a very wide companion HD 196982 (AT Mic), which is itself a binary system. \cite{Elliott2016} suggest a limit for wide binaries in the beta Pic moving group that agrees with HD 197481 (AU Mic) and HD 196982 (AT Mic) being bound in the past. However, the systems are considered comoving but not bound in \cite{Tokovinin2018}.}
    
\end{threeparttable}
\end{adjustbox}
\label{tab:companions}
\end{table*}

\end{document}